\documentstyle[twoside,oldlfont]{article}

\catcode`\@=11
\long\def\@makefntext#1{ 
\protect\noindent \hbox to 3.2pt {\hskip-.9pt
$^{{\eightrm\@thefnmark}}$\hfil}#1\hfill} 

\def\thefootnote{\fnsymbol{footnote}}
 \def\@makefnmark{\hbox to 0pt{$^{\@thefnmark}$\hss}}  

\def\ps@myheadings{\let\@mkboth\@gobbletwo
\def\@oddhead{\hbox{} 
\rightmark\hfil\eightrm\thepage}
\def\@oddfoot{}\def\@evenhead{\eightrm\thepage\hfil 
\leftmark\hbox{}}\def\@evenfoot{}
\def\sectionmark##1{}\def\subsectionmark##1{}}

\oddsidemargin=\evensidemargin
\renewcommand{\thefootnote}{\fnsymbol{footnote}}

\newcounter{sectionc}\newcounter{subsectionc}\newcounter{subsubsectionc}
\renewcommand{\section}[1] {\vspace{12pt}\addtocounter{sectionc}{1}
\setcounter{subsectionc}{0}\setcounter{subsubsectionc}{0}\noindent
	{\tenbf\thesectionc. #1}\par\vspace{5pt}}
\renewcommand{\subsection}[1] {\vspace{12pt}\addtocounter{subsectionc}{1}
	\setcounter{subsubsectionc}{0}\noindent
	{\bf\thesectionc.\thesubsectionc. {\kern1pt \bfit #1}}\par\vspace{5pt}}
\renewcommand{\subsubsection}[1] {\vspace{12pt}\addtocounter{subsubsectionc}{1}
	\noindent{\tenrm\thesectionc.\thesubsectionc.\thesubsubsectionc.
	{\kern1pt \tenit #1}}\par\vspace{5pt}}
\newcommand{\nonumsection}[1] {\vspace{12pt}\noindent{\tenbf #1}
	\par\vspace{5pt}}

\newcounter{appendixc}
\newcounter{subappendixc}[appendixc]
\newcounter{subsubappendixc}[subappendixc]
\renewcommand{\thesubappendixc}{\Alph{appendixc}.\arabic{subappendixc}}
\renewcommand{\thesubsubappendixc}
	{\Alph{appendixc}.\arabic{subappendixc}.\arabic{subsubappendixc}}

\renewcommand{\appendix}[1] {\vspace{12pt}
        \refstepcounter{appendixc}
        \setcounter{figure}{0}
        \setcounter{table}{0}
        \setcounter{lemma}{0}
        \setcounter{theorem}{0}
        \setcounter{corollary}{0}
        \setcounter{definition}{0}
        \setcounter{equation}{0}
        \renewcommand{\thefigure}{\Alph{appendixc}.\arabic{figure}}
        \renewcommand{\thetable}{\Alph{appendixc}.\arabic{table}}
        \renewcommand{\theappendixc}{\Alph{appendixc}}
        \renewcommand{\thelemma}{\Alph{appendixc}.\arabic{lemma}}
        \renewcommand{\thetheorem}{\Alph{appendixc}.\arabic{theorem}}
        \renewcommand{\thedefinition}{\Alph{appendixc}.\arabic{definition}}
        \renewcommand{\thecorollary}{\Alph{appendixc}.\arabic{corollary}}
        \renewcommand{\theequation}{\Alph{appendixc}.\arabic{equation}}
        \noindent{\tenbf Appendix \theappendixc #1}\par\vspace{5pt}}
\newcommand{\subappendix}[1] {\vspace{12pt}
        \refstepcounter{subappendixc}
        \noindent{\bf Appendix \thesubappendixc. {\kern1pt \bfit #1}}
	\par\vspace{5pt}}
\newcommand{\subsubappendix}[1] {\vspace{12pt}
        \refstepcounter{subsubappendixc}
        \noindent{\rm Appendix \thesubsubappendixc. {\kern1pt \tenit #1}}
	\par\vspace{5pt}}

\newcommand{\textlineskip}{\baselineskip=13pt}
\newcommand{\smalllineskip}{\baselineskip=10pt}

\def\eightcirc{
\begin{picture}(0,0)
\put(4.4,1.8){\circle{6.5}}
\end{picture}}
\def\eightcopyright{\eightcirc\kern2.7pt\hbox{\eightrm c}}

\def\abstracts#1#2#3{{
	\centering{\begin{minipage}{4.5in}\baselineskip=10pt\eightrm
	\centerline{ABSTRACT}
	\parindent=0pt #1\par
	\parindent=15pt #2\par
	\parindent=15pt #3
	\end{minipage} }\par}}

\renewenvironment{thebibliography}[1]			
	{\ninerm
	 \baselineskip=11pt				
	 \begin{list}{\arabic{enumi}.}
	{\usecounter{enumi}\setlength{\parsep}{0pt}
	 \setlength{\leftmargin 17pt}{\rightmargin 0pt}	
	 \setlength{\itemsep}{0pt} \settowidth		
	{\labelwidth}{#1.}\sloppy}}{\end{list}}

\newcounter{itemlistc}
\newcounter{romanlistc}
\newcounter{alphlistc}
\newcounter{arabiclistc}

\newcommand{\fcaption}[1]{
        \refstepcounter{figure}
        \setbox\@tempboxa = \hbox{\eightrm Fig.~\thefigure. #1}
        \ifdim \wd\@tempboxa > 5in
           {\begin{center}
        \parbox{5in}{\eightrm \smalllineskip Fig.~\thefigure. #1 }
            \end{center}}
        \else
             {\begin{center}
             {\eightrm Fig.~\thefigure. #1}
              \end{center}}
        \fi}

\newcommand{\tcaption}[1]{
        \refstepcounter{table}
        \setbox\@tempboxa = \hbox{\eightrm Table~\thetable. #1}
        \ifdim \wd\@tempboxa > 5in
           {\begin{center}
        \parbox{5in}{\eightrm\smalllineskip Table~\thetable. #1 }
            \end{center}}
        \else
             {\begin{center}
             {\eightrm Table~\thetable. #1}
              \end{center}}
        \fi}

\def\@citex[#1]#2{\if@filesw\immediate\write\@auxout	
	{\string\citation{#2}}\fi			
\def\@citea{}\@cite{\@for\@citeb:=#2\do			
	{\@citea\def\@citea{,}\@ifundefined		
	{b@\@citeb}{{\bf ?}\@warning
	{Citation `\@citeb' on page \thepage \space undefined}}
	{\csname b@\@citeb\endcsname}}}{#1}}

\newif\if@cghi
\def\cite{\@cghitrue\@ifnextchar [{\@tempswatrue
	\@citex}{\@tempswafalse\@citex[]}}
\def\citelow{\@cghifalse\@ifnextchar [{\@tempswatrue
	\@citex}{\@tempswafalse\@citex[]}}
\def\@cite#1#2{{$\null^{#1}$\if@tempswa\typeout
	{IJCGA warning: optional citation argument
	ignored: `#2'} \fi}}

\def\pmb#1{\setbox0=\hbox{#1}
	\kern-.025em\copy0\kern-\wd0
	\kern.05em\copy0\kern-\wd0
	\kern-.025em\raise.0433em\box0}

\def\fnt#1#2{\footnotetext{\kern-.3em
	{$^{\mbox{\scriptsize #1}}$}{#2}}}

\def\fpage#1{\begingroup
\voffset=.3in
\thispagestyle{empty}\begin{table}[b]\centerline{\footnotesize #1}
	\end{table}\endgroup}

\def\runninghead#1#2{\pagestyle{myheadings}
\markboth{{\eightit{\quad #1}}\hfill}{\hfill{\eightit{#2\quad}}}}

\font\tenbf=cmbx10
\font\tenit=cmti10
\font\tenit=cmti10
\font\bfit=cmbxti10 at 10pt
 1
 1
 1

\font\ninerm=cmr9

\font\eightrm=cmr8
\font\eightit=cmti8

\def\qed{\hbox{${\vcenter{\vbox{                          
   \hrule height 0.4pt\hbox{\vrule width 0.4pt height 6pt
   \kern5pt\vrule width 0.4pt}\hrule height 0.4pt}}}$}}

\runninghead{}{}

\newcommand{\nn}{\noindent}
\newcommand{\non}{\nonumber}
\newcommand{\s}{\\ \vspace*{-1mm}}
\newcommand{\ee}{e^+e^-}
\newcommand{\ra}{\rightarrow }
\newcommand{\lra}{\longrightarrow }
\newcommand{\SM}{{\cal SM}}
\newcommand{\MSSM}{{\cal MSSM}}
\newcommand{\SUSY}{{\cal SUSY}}
\newcommand{\CP}{{\cal CP}}
\newcommand{\beq}{\begin{eqnarray}}
\newcommand{\eeq}{\end{eqnarray}}
\newcommand{\ga}{\gamma \gamma}
\newcommand{\tb}{{\rm tg} \beta}
\newcommand{\tg}{{\rm tg} \beta}
\newcommand{\X}{{\cal H}}

\begin{document}
\normalsize\textlineskip
{\thispagestyle{empty}
\setcounter{page}{1}

\renewcommand{\thefootnote}{\fnsymbol{footnote}} 

\vspace*{-1.5cm}

\nn \hspace*{9cm} UdeM-GPP-TH-94-01 \\
    \hspace*{9cm} PM 94/27 \\

\fpage{1}
\centerline{\bf HIGGS PARTICLES AT FUTURE}
\vspace*{0.035truein}
\centerline{\bf HADRON AND ELECTRON--POSITRON COLLIDERS\footnote{\normalsize
Invited Review Paper for International Journal of Modern Physics.}}
\vspace*{0.15truein}
\centerline{\footnotesize A. DJOUADI\footnote{\normalsize NSERC Fellow}}
\vspace*{0.015truein}
\centerline{\footnotesize\it Groupe de Physique des Particules, Universit\'e
de Montr\'eal, Case 6128A,}
\baselineskip=10pt
\centerline{\footnotesize\it H3C 3J7 Montr\'eal PQ, Canada\footnote{
\normalsize Present adress.}}
\baselineskip=10pt
\centerline{\footnotesize\it and}
\baselineskip=10pt
\centerline{\footnotesize\it Physique Math\'ematique et Th\'eorique,
U.R.A. 768 du CNRS,}
\baselineskip=10pt
\centerline{\footnotesize\it Universit\'e de Montpellier II, 34095 Montpellier
Cedex 5, France}
\vglue 10pt

\vspace*{0.15truein}

\abstracts{\nn The prospects for discovering Higgs particles and studying
their fundamental properties at future high--energy electron--positron and
hadron colliders are reviewed. Both the Standard Model Higgs boson
and the Higgs particles of its minimal supersymmetric extension are discussed.
We update various results by taking into account the new value of the top
quark mass obtained by the CDF Collaboration and by including radiative
corrections some of which have been calculated only recently.}{}{}

\vspace*{3mm}
\vspace*{-5pt}\textlineskip

\renewcommand{\theequation}{1.\arabic{equation}}
\setcounter{equation}{0}

\section{Introduction}

\vspace*{-3mm}

\subsection{Standard Model Higgs Sector}

\nn The Higgs mechanism\cite{R1} is a cornerstone in the electroweak sector
of the Standard Model ($\SM$).\cite{R2} The fundamental particles, leptons,
quarks and weak gauge bosons, acquire their masses through the interaction with
a scalar field. The self--interaction of the scalar field leads to a non--zero
field strength in the ground state, inducing the spontaneous breaking of the
electroweak SU(2)$\times$U(1) symmetry down to the electromagnetic U(1)
symmetry.

In order to accomodate the well--established electromagnetic and weak
phenomena, this mechanism for generating particle masses requires the existence
of at least one weak iso--doublet scalar field. The three Goldstone bosons
among the four degrees of freedom are absorbed to build up the longitudinal
polarization states of the massive $W^\pm,Z$ gauge bosons. One degree of
freedom is left over, corresponding to a real physical scalar particle. The
discovery of this Higgs particle is the {\it experimentum crucis} for the
standard formulation of the electroweak interactions.

The only unknown parameter in the $\SM$ Higgs sector is the mass of the
Higgs particle. Even though the value of the Higgs mass cannot be predicted
in the Standard Model, interesting constraints can be derived from assumptions
on the energy range within which the model be valid before perturbation
theory breaks down at a scale $\Lambda$ and new dynamical phenomena would
emerge.

\newpage

$(i)$ If the Higgs mass is larger than $\sim$ 1 TeV, the interactions between
longitudinal $W$ and $Z$ bosons would become so strong that non--perturbative
effects are needed to ensure unitarity at high energies.\cite{R3} In the 500
GeV energy range, residual final state interactions in $e^+e^- \rightarrow
W^+W^-$ would still be too small to be observable; their detection would
require much higher energy $\ee$ machines. However, this scenario can
eventually be studied at proton colliders in an exploratory way.

$(ii)$ The strength of the Higgs self--interaction is determined by the
Higgs mass itself at the scale $M_H$. Increasing the scale, the quartic
self--coupling of the Higgs field grows logarithmically with the energy
scale,\cite{R4,R5} similarly to the electromagnetic coupling in QED.
If the Higgs
mass is small, the energy cut--off $\Lambda$ is large at which the coupling
grows beyond any bound and new phenomena may be observed; conversely, if the
Higgs mass is large, the cut--off $\Lambda$ is small. The condition $M_H <
\Lambda$ sets an upper limit on the Higgs mass in the Standard Model. Thorough
analyses of the non--perturbative regime near the cut--off lead to an estimate
of about 630 GeV for the upper limit of $M_H$.\cite{R50} However, if the Higgs
mass is less than 180 to 200 GeV, the Standard Model can be extended up to the
GUT scale $\Lambda_{{\rm GUT}} \sim 10^{15}$ GeV with weakly interacting
particles. Including the effect of $t$--quark loops on the running coupling, a
detailed analysis predicts the area of the allowed $(m_t, M_H)$ values shown in
Fig.~1 for several values of the cut--off parameter $\Lambda$.\cite{R5}

On quite general grounds, the hypothesis that interactions of $W,Z$ bosons
and Higgs particles remain weak up to the GUT scale is an attractive idea and
may play a key r\^ole in the explanation of the experimental value of the
electroweak mixing parameter $\sin^2 \theta_W$. Based on the $\SM$ particle
spectrum, the mixing parameter evolves from the symmetry value $3/8$ at the GUT
scale down to $\sim 0.2$ at ${\cal O}(100$ GeV).\cite{R5A} Even though
additional degrees of freedom are needed\cite{R5B} to account for the small
discrepancy to the experimentally observed value $0.23$, the hypothesis that
the particle interactions remain weak up to the GUT scale is nevertheless
qualitatively supported by this result.

$(iii)$ Top--quark quantum corrections to the quartic Higgs coupling are
negative, driving the coupling to negative values for which the vacuum becomes
instable. The boundary on the right hand side  of the allowed domain in the
$(m_t,M_H)$ plane corresponds to the values where the quartic coupling
vanishes. For top masses larger than about 100 GeV, this leads to a lower limit
on the Higgs mass. [The $(m_t,M_H)$ mass pairs are attracted by the line
connecting the tips of the allowed areas if the theory evolves from the
cut--off energies $\Lambda$ down to the ${\cal O}$(100 GeV) range.\cite{R6}]

{}From the requirement of vacuum stability and from the assumption that the
Standard Model can be continued up to the GUT scale, upper and lower bounds on
the Higgs mass can be derived. Based on these arguments, the Higgs mass could
well be expected in the window $100 < M_H < 180$ GeV for a top mass value of
150 GeV. It must however be stressed again that the upper limit is based on
assumptions which are backed only {\it qualitatively} by the measured value of
$\sin^2\theta_W$.

The important range of $M_H$ between $M_Z$ and $2M_Z \simeq 180$ GeV is
generally referred to as the intermediate Higgs mass range.

\begin{small}

\nn \hspace*{8cm} \begin{minipage}[r]{45mm}{Fig.~1 Allowed values of
the top and Higgs masses if the Standard Model can be extended up to the scale
$\Lambda$; from Ref.\cite{R5} The dashed area follows from the assumption
that the Standard Model can be extended up to the GUT scale. The lower bound is
derived from vacuum stability.} \end{minipage}

\end{small}

\vspace*{4.5cm}

Once the Higgs mass is fixed, the triple and quartic self--couplings of the
$\SM$ Higgs particle are uniquely determined [at the tree level]. The size of
the Higgs couplings to massive gauge bosons and quarks/leptons is set by the
masses of these particles. Hence, the profile of the Higgs particle can be
predicted completely for a given value of the Higgs mass: the decay properties
are fixed  and the production mechanisms and the production rates can be
determined.

So far, the most comprehensive search for Higgs particles has been carried out
in $Z$ decays at LEP,\cite{R7,R8} based on the Bjorken process $Z \rightarrow
Z^*H$. By now, a lower bound on the Higgs mass of about $M_H \geq 63.8$ GeV
could be established.\cite{AX1} This limit can be raised by a few GeV by
accumulating more statistics. In the second phase of LEP with a total energy
close to 200 GeV, Higgs particles can be searched for up to masses of ~85 to 90
GeV in Higgs bremsstrahlung off the $Z$ line.\cite{R8A} Higher energy colliders
are required to sweep the entire mass range for the Higgs particle.

\subsection{Supersymmetric Extension}

\nn Supersymmetric theories ($\SUSY$) are very attractive extensions of the
Standard Model, incorporating the most general symmetry of the ${\cal S}$
matrix in field theory.\cite{Haag} At low energies they provide a theoretical
framework in
which the problem of naturalness and hierarchy in the Higgs sector is solved
by retaining Higgs bosons with moderate masses as elementary particles in
the presence of high mass scales demanded by grand unification. The Minimal
Supersymmetric extension of the Standard Model\cite{S1,S2} ($\MSSM$) may serve
as a useful guideline into this domain. This point is underlined by the fact
that the model led to a prediction of the electroweak mixing angle\cite{R5B}
that is in very nice agreement with present high--precision measurements of
$\sin^2\theta_W$. Although some of the phenomena will be specific to this
minimal version, the general pattern will nevertheless be characteristic to
more general extensions\cite{S2,S3,S4} so that the $\MSSM$ can be considered as
representative for a wide class of $\SUSY$ models.

Supersymmetry requires the existence of at least two isodoublet fields
$\Phi_{1}$ and $\Phi_{2}$, thus extending the physical spectrum of
scalar particles  to five. The $\MSSM$ is restricted to this minimal extension.
The field $\Phi_{2}$ [with vacuum expectation value $v_{2}$] couples only to
up--type quarks while $\Phi_{1}$ [with vacuum expectation value $v_{1}$]
couples to down--type quarks and charged leptons. The physical Higgs bosons
introduced by this extension are of the following type: two ${\cal CP}$--even
neutral bosons $h$ and $H$ [where $h$ will be the lighter particle], a ${\cal
CP}$--odd neutral boson $A$ [usually called pseudoscalar] and two charged Higgs
bosons $H^{\pm}$. Besides the four masses $M_h$, $M_H$, $M_A$ and $M_{H^\pm}$,
two additional parameters define the properties of the scalar particles and
their interactions with gauge bosons and fermions:	the ratio of the two
vacuum expectation values $\tb = v_2/v_1$ and a mixing angle $\alpha$ in the
neutral ${\cal CP}$--even sector. Supersymmetry leads to several relations
among these parameters and, in fact, only two of them are independent at tree
level.  These relations impose a strong hierarchical structure on the mass
spectrum $ [ M_h<M_Z , M_A < M_H$ and $M_W <M_{H^\pm}]$  which however is
broken by radiative corrections\cite{S5} since the top quark mass is large. The
parameter $\tg$ will in general be assumed in the range $1 < \tg < m_t/m_b$ $[
\pi/4 < \beta < \pi/2] $, consistent with restrictions\cite{S5a} that follow
from interpreting the $\MSSM$ as low energy limit of a supergravity model.

Since the lightest ${\cal CP}$--even scalar boson $h$ is likely to be the Higgs
particle which will be discovered first, an attractive choice of the two input
parameters is the set $[M_h, \tb $], with $\tb$ to be determined by the
production cross sections. Once these two parameters [as well as the top quark
mass and the associated squark masses which enter through radiative
corrections] are specified, all other masses and the mixing angle $\alpha$ can
be predicted. To incorporate radiative corrections we first shall neglect, for
the sake of simplicity, non--leading effects due to non--zero values of the
supersymmetric Higgs mass parameter $\mu$ and	of the parameters $A_t$ and
$A_b$ in the soft symmetry breaking interaction. The radiative corrections are
then determined by the parameter $\epsilon$ which grows as the fourth power of
the top quark mass $m_t$ and logarithmically with the squark mass
$M_S$\cite{S5}
\begin{eqnarray}
\epsilon = \frac{3 \alpha}{2 \pi} \frac{1}{s_W^2 c_W^2}
\frac{1}{\sin^2	\beta} \frac{m_t^4}{M_Z^2} \log	\left( 1+
\frac{M_S^2}{m_t^2} \right)
\end{eqnarray}
[$s_W^2=1-c_W^2 \equiv \sin^2 \theta_W$].  These corrections are positive
and they shift the mass of	the light neutral Higgs boson $h$ upward with
increasing top mass.	The variation of the upper limit on $M_h$ with
the	top quark mass is shown	in Fig.~2 for $M_S=$ 1 TeV and two
representative values of $\tb =2.5$ and 20; and update of Ref.\cite{X5}
While the
dashed lines correspond to the leading radiative corrections in eq.~(1.1)
[$\mu=A_t=A_b=0$], the solid lines correspond to $\mu=-200,0,+200$ GeV and
$A_t=A_b=1$ TeV. The upper bound on $M_h$ is shifted from the tree level value
$M_Z$ up to $\sim $ 130 GeV for $m_t=175$ GeV and $\sim$ 140 GeV for $m_t=200$
GeV.

\newpage

\vspace*{16.9cm}

\nn {\small Fig.~2 The masses of the Higgs particles in the $\MSSM$ including
radiative corrections; the squark masses are fixed to 1 TeV. The dashed curve
shows the leading correction [$A_t=A_b=\mu=0]$ while the solid curves include
the full corrections [$A_t=A_b=1$~TeV and $\mu=-200,0,200$~GeV] (a) Upper
limit on $M_h$ as a function of $m_t$; (b)--(d) masses of the $H,A$ and $H^\pm$
Higgs bosons as functions of $M_h$. The top mass is fixed to 175 GeV. }

\newpage

Taking $M_h$ and $\tb$ as the input parameters, the mass of the pseudoscalar
$A$  is given by
\begin{eqnarray}
M_A^2= \frac{M_h^2(M_Z^2-M_h^2+\epsilon)-\epsilon M_Z^2 \cos^2 \beta}
{M_Z^2 \cos^2 2\beta -M_h^2+ \epsilon \sin^2 \beta}
\end{eqnarray}
\nn and the masses of the heavy neutral and charged Higgs bosons follow from
the sum rules
\begin{eqnarray}
M_H^2 &	= & M_A^2+M_Z^2-M_h^2+\epsilon	\\
M_{H^\pm}^2 & =	& M_{A}^{2}+M_{W}^{2}
\end{eqnarray}
In the subsequent discussion, we will use for definiteness the two values
$m_t=175$ GeV and $M_S=1$ TeV. For the two representative values of $\tb$
introduced above, the masses $M_A,M_H$ and $M_{H^\pm}$ are displayed 	in
Fig.~2b--d as a function of the light neutral Higgs mass $M_h$.
Apart	from the range near the upper limit of $M_h$ for a given value of
$\tb$, the masses cluster in characteristic bands of 100 to 200 GeV for
$M_H$ and $M_{H^\pm}$, and up to $\sim	$ 150 GeV for $M_A$ [similarly
to $M_h$]. On general grounds, the masses of the heavy neutral and charged
Higgs bosons are expected to be of the order of the electroweak symmetry
breaking scale.

The	mixing parameter $\alpha$ is determined by $\tb$ and the Higgs masses,
\begin{eqnarray}
{\rm tg} 2 \alpha = {\rm tg} 2 \beta \frac{M_{A}^{2} +M_{Z}^{2}}{M_{A}^{2}-
M_{Z}^{2}+ \epsilon /\cos 2\beta} \hspace*{2cm} \left[ \ -\frac{\pi}{2} \leq
\alpha \leq 0 \ \right]
\end{eqnarray}

The couplings of the various neutral Higgs bosons to fermions and gauge
bosons will in general depend	on the angles $\alpha$ and $\beta$. Normalized
to the $\SM$ Higgs couplings, they are summarized in Table 1.  The pseudoscalar
particle $A$ has no tree level couplings to gauge bosons, and its couplings to
down (up) type fermions are (inversely) proportional to $\tb$.

\vspace*{0.5cm}

\nn {\small Tab.~1: Higgs bosons couplings in the $\MSSM$ to fermions and gauge
bosons relative to the $\SM$ Higgs boson couplings.} \\

\begin{center}
\begin{tabular}{|c||c|c||c|c|} \hline
& & & \\
$\hspace{1cm} \Phi \hspace{1cm} $ &$ g_{ \Phi \bar{u} u} $ & $ g_{\Phi \bar{d}
d} $ & $g_{ \Phi VV} $ \\
& & & \\ \hline \hline
& & & \\
$H_{SM}$ & \ $ \; 1  \;	$ \ & \	$ \; 1  \; $ \	& \ $ \; 1  \; $ \ \\
$h$  & \ $\; \cos\alpha/\sin\beta	\; $ \ & \ $ \;	-\sin\alpha/
\cos\beta \; $ \ & \ $ \; \sin(\beta-\alpha) \;	$ \ \\
 $H$  & \	$\; \sin\alpha/\sin\beta \; $ \	& \ $ \; \cos\alpha/
\cos\beta \; $ \ & \ $ \; \cos(\beta-\alpha) \;	$ \ \\
$A$  & \ $\; 1/ \tg \; $	\ & \ $	\; \tg \; $ \
& \ $ \; 0 \; $	\ \\
& & & \\ \hline
\end{tabular}
\end{center}

\newpage

\vspace*{18cm}

\nn {\small Fig.~3 Coupling parameters of $\SUSY$ Higgs bosons as functions of
the lightest Higgs boson mass $M_h$. The couplings are normalized to the $\SM$
Higgs boson couplings as in Tab.~1. The parameters are the same as in Fig.~2.}

\newpage

Typical numerical values of	these couplings	are shown in Fig.~3 as a
function of the lightest neutral Higgs boson mass and for two values of $\tb$;
the same set of $\mu$ and $A_t,A_b$ values as in Fig.~2 has been chosen. The
figure demonstrates that the dependence on the parameters $\mu$ and $A_t,A_b$
is very weak and the leading radiative corrections provide a very good
approximation. [This is also the case for the radiative corrections to the
Higgs boson masses, Fig.~2, although the dependence on $A_t,A_b$ is slightly
stronger in this case, leading to a shift of a few GeV]. There is in general a
strong dependence on the input parameters $\tb$ and $M_h$.  The
couplings	to down	(up) type fermions are enhanced
(suppressed)	compared to the	$\SM$ Higgs couplings.  If $M_h$ is	very
close to its upper limit for a given value of $\tb$,	the couplings of $h$
to fermions and gauge bosons are $\SM$ like.  It is therefore very difficult to
distinguish the Higgs	sector of the $\MSSM$ from the $\SM$, if all other
Higgs bosons are very heavy.

Negative searches for $\SUSY$ Higgs particles at LEP in the processes $Z
\rightarrow Z^*H$ and $Z \rightarrow AH$ exclude\cite{LEPSUSY}  $h$ and $A$
bosons with masses smaller than $M_{h,A} \simeq 45$ GeV for $m_t=140$ GeV,
$M_S=1$ TeV and tg$\beta >1$ [if the parameters of the model are allowed to
vary arbitrarily and if one includes all possible decay modes, this bound
becomes\cite{LEPSUSY} $M_h >44$ GeV and $M_A>21$ GeV].

Charginos and neutralinos are expected to be the lightest supersymmetric
particles. In general, they are mixtures of the [non--colored] gauginos and
Higgsinos, the spin 1/2 supersymmetric partners of the gauge bosons and Higgs
bosons. There are two chargino $\tilde{\chi}_i^\pm$ [$i=1,2$] and four
neutralino $\tilde{\chi}_i^0$ [$i=1,..,4$] states, where $\tilde{\chi}_1^0$
will be assumed to be the stable lightest supersymmetric particle. The masses
and the couplings of these particles are obtained by diagonalizing the charged
and neutral mass matrices; the unitary matrices which diagonalize the mass
matrices can be found in Ref.\cite{S6} Interpreting
the $\MSSM$ as the low--energy
limit of a supergravity model, the matrix elements will depend on four
parameters, one Higgs mass [$M_h$ or $M_A$], $\tb, \mu$ and the SU(2) gaugino
mass $M$ which, without loss of generality, can be taken to be positive. [The
fifth $\SUSY$ parameter of the model is the universal mass parameter $m_0$ of
the scalar particles at the unification scale.]

{}From the negative search of supersymmetric particles in $Z$ decays,\cite{R8}
the lightest neutralino [$\tilde{\chi}_1^0$] mass is restricted to be larger
than 20 GeV for $\tb=2.5$ and larger than 22 GeV for $\tb >4$; the second
lightest neutralino [$\tilde{\chi }_2^0$] and the charginos are excluded if
their masses are less than $ \sim M_Z/2$. If the search at LEP200 with a c.m.
energy of 180 GeV is negative, charginos with masses $m_{\tilde{\chi}^+_1}<90$
GeV will also be excluded. The present LEP data also exclude sleptons with
masses below $\sim M_Z/2$; if sleptons will not be observed at LEP200  these
limits can be improved by roughly a factor of two. On the other hand, CDF
data\cite{CDFSQUARK} restrict the squarks masses to be larger than $\sim$ 150
GeV if cascade decays are suppressed. Since the sfermion masses depend on the
universal mass parameter $m_0$, they are not fixed by the mass scales which
appear in the chargino/neutralino mass matrices. We shall assume in this
discussion that squarks and sleptons are heavy so that they will not affect
Higgs boson decays and production.

\subsection{Organization of the paper}

\nn The discovery of Higgs particles is the most fundamental test of the modern
formulation of the electroweak theory, and therefore is the major goal of
future accelerators. In this review, we will discuss the prospects for
producing Higgs particles [and once produced, for studying their fundamental
properties] in the Standard Model as well as in its minimal supersymmetric
extension at future high--energy colliders: the CERN proton--proton collider
LHC with a center of mass energy of $\sim 14$ TeV and a future $\ee$ linear
collider [which can also be turned into to a high--energy $\ga$ collider] with
an energy in the range 300 to 500 GeV.

We will update various results for Higgs masses and couplings [in the
supersymmetric extension of the Standard Model], partial decay widths and
production cross sections, to take into account the value of the top quark mass
recently published by the CDF collaboration\cite{CDFTOP}
\begin{eqnarray}
m_t=174 \ ^{+10}_{-11} \ \pm \ 10 \ {\rm GeV}
\end{eqnarray}
with a central value which coincides with the favored value obtained from
a global fit\cite{LEPTOP} of electroweak precision measurements at LEP and SLC
\begin{eqnarray}
m_t=174 \ ^{+11}_{-12} \ ^{+17}_{-19} \ {\rm GeV}
\end{eqnarray}

Whenever possible we will use the value $m_t=175$ GeV, although sometimes
we will also use the values 150 or 200 GeV which can be viewed as conservative
lower and upper bounds on the top mass, respectively.

In section 2, we will summarize the various decay modes of the Higgs particles
including a discussion of the main QCD and electroweak radiative corrections,
some of which have been calculated only recently. Partial and total decay
widths
as well as the most important decay branching fractions will be given. In the
case of supersymmetric Higgs bosons, both the standard and the supersymmetric
decay modes will be discussed.

In section 3, we will discuss Higgs production at the CERN ``Large Hadron
Collider" (LHC).\cite{LHC} We will summarize the various production processes,
with a special emphasis on the dominant one, gluon--gluon fusion via a heavy
quark loop, for which we will discuss the QCD corrections [which turned out to
be very important] and the leading electroweak correction. The main detection
channels for standard and supersymmetric Higgs bosons will be summarized.

In section 4, we will analyze the potential of a future $\ee$ linear
collider with a center of mass energy of 500 GeV.\cite{EE1,EE2,EE2C,EE3} We
will
discuss in some detail the main production processes, especially the
bremsstrahlung process, and the corresponding backgrounds. We then analyze the
possibility of measuring some fundamental properties of the Higgs particles,
such as their couplings to the other elementary particles and their
spin--parity quantum number assignements. A short discussion about the
complementary potential of $\ga$ colliders will be given.

Section 5 contains our conclusions.


\renewcommand{\theequation}{2.\arabic{equation}}
\setcounter{equation}{0}

\section{Decays Modes}

\vspace*{-3mm}

\subsection{Standard Model}

\nn As previously discussed, the profile of the Higgs particle in the Standard
Model is uniquely determined at tree level if the Higgs mass is fixed.  The
strength of the Yukawa couplings of the Higgs boson to fermions is set by the
fermion masses $m_f$, and the coupling to the electroweak gauge bosons $V=W,Z$
by their masses $M_V$:
\beq
g_{ffH} & =& \left[ \sqrt{2} G_F \right]^{1/2} m_f \\
g_{VVH} & =& 2 \left[ \sqrt{2} G_F \right]^{1/2} M_V
\eeq

The decay width, the branching ratios and the production cross sections
are given by these parameters. The partial widths for the main decay channels,
including the dominant QCD and electroweak radiative corrections, and their
characteristics are summarized below; an update of Ref.\cite{PR2}

In the Born approximation the width of the Higgs decay into
lepton pairs is\cite{V2,V3}
\beq
\Gamma (H \ra l^+ l^-)= \frac{G_F m_l^2}{4 \sqrt{2} \pi} \ M_H \ \beta^3
\eeq
with $\beta=(1- 4m_l^2/M_H^2)^{1/2}$ being the velocity of the leptons in
the final state. For the decay widths into quark pairs, eq.~(2.3) has to be
supplemented by a color factor $N_c=3$ in the Born approximation. However, the
QCD corrections\cite{V4,V4A,DG} turn out to be quite large and therefore must
be included, and the partial decay width  reads
\beq
\Gamma (H \ra q \bar{q} )= \frac{3G_F m_q^2}{4 \sqrt{2} \pi} \ M_H \ \beta^3
\left[ 1 +\frac{4}{3} \frac{\alpha_S}{\pi} \Delta_{H}^{QCD} \right]
\eeq
where the QCD correction factor is given by\cite{V4,V4A,DG}
\beq
\Delta_{H}^{QCD} = \frac{1}{\beta}A(\beta) + \frac{1}{16\beta^3}(3+34\beta^2-
13 \beta^4)\log \frac{1+\beta}{1-\beta} +\frac{3}{8\beta^2}(7 \beta^2-1)
\eeq
with [Li$_2$ is the Spence function defined by, Li$_2(x)= -\int_0^x dy y^{-1}
\log(1-y)$]
\beq
A(\beta) &= & (1+\beta^2) \left[ 4 {\rm Li}_2 \left( \frac{1-\beta}{1+\beta}
\right) +2 {\rm Li}_2 \left( -\frac{1-\beta}{1+\beta} \right) -3 \log
\frac{1+\beta}{1-\beta} \log \frac{2}{1+\beta} \right. \non \\
& & \left. -2 \log \frac{1+\beta}{1-\beta} \log \beta \right] -
3 \beta \log \frac{4}{1-\beta^2} -4 \beta \log \beta
\eeq
In the limit $M_H \gg m_q$, the decay width receives logarithmic contributions
which in the case of the $b$ quark and for Higgs masses around 100 GeV,
decrease the $ H \ra b\bar{b}$ decay width by more than 50\%. The bulk of these
QCD corrections can be absorbed into running quark masses evaluated at the
scale $\mu=M_H$.\cite{V4,V5} In the limit $M_H \gg m_q$, one can also include
the order $\alpha_s^2$ corrections which are known,\cite{V5} and the decay
width reads
\beq
\Gamma (H \ra q \bar{q}) = \frac{3 G_F}{ 4\sqrt{2} \pi} m_q^2(M_H^2)
M_H \left[ 1+5.67 \left( \frac{\alpha_s}{\pi} \right)+(35.94-1.36N_F)
\left( \frac{\alpha_s}{\pi} \right)^2 \right]
\eeq
$\alpha_s \equiv \alpha_s(M_H^2)$ and $N_F=5$ is the number of active quark
flavors; all quantities are defined in the $\overline{{\rm MS}}$ scheme with
$\Lambda_{\overline{{\rm MS}}} \sim 150$ MeV for $N_F=5$. For $M_H=120$ GeV,
the $b$ and $c$ quark masses $m_b(m_b^2)=4.2$ GeV and $m_c(m_c^2)=1.35$ GeV
have dropped to the effective values $m_b(M_H^2)=3$ GeV and $m_c(M_H^2)=0.77$
GeV, respectively. For Higgs decays to top quarks, the QCD corrections are not
large [except near threshold] since $m_t$ is of the same order as $M_H$ and
the decay width is approximately given by eq.~(2.3) apart from the additional
color factor $N_c=3$.

The electroweak radiative corrections to fermionic Higgs decays are well
under control.\cite{V6,V7} Parametrizing the Born formulae (2.3) in terms of
the Fermi coupling constant $G_F$, the corrections are free of large logarithms
associated with light fermion loops. Below the $W$ threshold, the total
correction $\delta=\delta_{EW}+\delta_{m}$ can be approximated by\cite{PR2}
[similarly to the QCD corrections, the additional logarithmic term
$\delta_m$ for photonic corrections can be mapped into the running mass]
\beq
\delta_{EW}=\frac{\alpha}{\pi} \left[ \frac{9}{4}Q_f^2+ \frac{1}{16s_W^2}
\left(k_f \frac{m_t^2}{M_W^2}-5 +\frac{3}{s_W^2} \log c_W^2 \right)- 3
\hat{v}_f^2 + \frac{1}{2}\hat{a}_f^2 \right]
\eeq
with $\hat{v_f}= (2 I^{3}_f-4 s_W^2 Q_f)/(4s_Wc_W)$ and $\hat{a_f}= 2 I^{3}_f
/(4s_Wc_W)$.
The coefficient $k_f=7$ for $\tau,c$ is reduced to $k_f=1$ for $b$ quarks and
the total correction for these fermions is of the order of a few percent.

Above the $H \ra WW$ and $ZZ$ decay thresholds, the partial width into
massive gauge boson pairs may be written as\cite{R3}
\begin{eqnarray}
\Gamma (H \ra VV) = \delta_V \frac{\sqrt{2}G_F}{32 \pi} M_H^3 (1-4x+12x^2)
\beta\,,
\end{eqnarray}
where $x=M_V^2/M_H^2$, $\beta=\sqrt{1-4x}\,$ and $\delta_V=2(1)$ for $V=W(Z)$.
For large Higgs masses, the vector bosons are longitudinally polarized. Since
the longitudinal wave functions are linear in the energy, the width grows as
the third power of the Higgs mass. The electroweak radiative
corrections\cite{V6,V8} are positive and amount to a few percent above the
threshold.

Below the threshold for two real bosons, the Higgs particle can decay into
real and virtual $VV^*$ pairs, primarily $WW^*$ pairs above $M_H \sim 110$ GeV.
The partial decay width, $W$ charges summed over, is given by\cite{V9}
\begin{eqnarray}
\Gamma (H \ra VV^*) = \frac{3 G_F^2 M_V^4}{16 \pi^3} M_H R(x) \delta_V'\,,
\end{eqnarray}
with $\delta'_W=1$ and $\delta_Z' =7/12-10\sin^2\theta_W/9+40\sin^4\theta_W/27$
and
\begin{eqnarray}
R(x) & = & \frac{3(1-8x+20x^2)}{(4x-1)^{1/2}} \arccos \left( \frac{3x-1}
{2x^{3/2}} \right) -\frac{1-x}{2x} (2-13x+47x^2) \non \\
&& - \frac{3}{2}(1-6x+4x^2) \log x\,.
\end{eqnarray}
The $\ga$\cite{V3,LET}, $\gamma Z$\cite{V11a} and $gg$\cite{V11} couplings to
Higgs bosons are mediated by heavy particle triangular loops. These loop decays
are important only for Higgs masses below $\sim 140$ GeV where the total decay
width is rather small. However, they are very interesting since their strength
is sensitive to scales far beyond the Higgs mass and can be used as a possible
telescope for new particles whose masses are generated by the Higgs mechanism.

In the Standard Model only the $W$ and top quark loops contribute significantly
to the $H \ga$ coupling. For a Higgs boson in the intermediate mass range, $M_W
\leq M_H \leq 2M_W$, the $W$ loop contribution is always dominating and
interferes destructively with the fermion amplitude. The decay width varies
from $\sim 5$ to $\sim 50 $ keV in this mass range. The QCD corrections to the
quark amplitude have been calculated in Refs.\cite{X1,V10,Melnikov} In the
intermediate mass range, the correction factor is well under control being of
${\cal O} (\alpha_s/\pi)$. Thus, contrary to the $H \ra \bar{q}q$ case, the QCD
corrections to $H \ra \ga$ do not generate large logarithms. A more detailed
discussion of these corrections will be given in section 4.3.

Similar to the $\ga$ case, the $H\ra Z\gamma$ coupling is built up
by top quark and $W$ loops; the $W$ loop being by far dominating. The decay
occurs for $M_H > M_Z$  and the decay width varies from a few keV for $M_H
\sim 120$ to $\sim 100$ keV for $M_H \sim 2M_W$. The QCD corrections to
the quark loop, calculated in Ref.\cite{X2}, are rather small in the previous
range for $M_H$, being of ${\cal O}(\alpha_s/\pi)$.

Gluonic Higgs decays, $H\ra gg$, are mediated by top quark loops\cite{V11}
in the Standard Model and the decay width is of significance only for Higgs
masses below the top threshold. Incorportaing the QCD radiative corrections
which include $ggg$ and $gq \bar{q}$ final states and which are very important
since they increase the partial width by $\sim 65\%$, the partial width can be
cast in the approximate form\cite{X3}
\beq
\Gamma( H\ra gg) = \frac{G_F \alpha_s^2(M_H^2)}{36 \sqrt{2} \pi^3} \ M_H^3 \
\left[ 1+ \frac{215}{12} \frac{\alpha_s (M_H^2)}
{\pi} \right]
\eeq

By adding up all possible decay channels, we obtain the total width shown
in Fig.~4a for $m_t=175$ GeV. Up to masses of 140 GeV, the Higgs particle is
very narrow, $\Gamma(H) \leq 10$ MeV. After opening the [virtual] gauge boson
channels, the state becomes rapidly wider, reaching $\sim$ 1 GeV at the $ZZ$
threshold. The width cannot be measured directly in the intermediate mass
range. Only above $M_H \geq 250$ GeV it becomes wide enough to be resolved
experimentally.

The branching ratios of the main decay modes are displayed in Fig.~4b; an
update of Ref.\cite{R9a} A large variety of channels will be accessible for
Higgs masses below 140 GeV. The by far dominant mode are $b \bar{b}$ decays,
yet $c \bar{c}$, $\tau^+ \tau^-$ and $gg$ still occur at a level of several
percent. [At $M_H=120$ GeV for instance , the branching ratios are 68\% for
$b\bar{b}$, 4.6\% for $c \bar{c}$, 6.6\% for $\tau^+ \tau^-$ and 6\% for $gg$.]
The branching ratios for the $H \ra \ga$ and $\gamma Z$ are small, being of
${\cal O}(10^{-3})$.  Above the mass value $M_H=140$ GeV, the Higgs boson decay
into $W$'s becomes dominant, overwhelming all other channels once the decay
mode into two real $W$'s is kinematically possible.

\newpage

\vspace*{18cm}

\nn {\small Fig.~4 Total decay width (a) and decay branching ratios (b) of the
$\SM$ Higgs boson; the top quark mass is fixed to $m_t=175$ GeV. The QCD
corrections to the hadronic decay modes are included.}

\subsection{Supersymmetric Extension}

\nn The lightest Higgs boson will decay mainly into fermion pairs since its
mass is smaller than $\sim$ 130 GeV. This is also the dominant decay mode
of the pseudoscalar boson $A$ which has no tree level couplings to gauge
bosons.	The partial decay width of a neutral Higgs boson $\Phi$ into fermion
pairs is given by
\begin{eqnarray}
\Gamma(\Phi \ra \bar{f} f) & = & N_{c} \frac{G_F m_f^2}{4\sqrt{2} \pi}
\ g_{\Phi ff}^2 \ M_{\Phi} \ \beta^{p}
\end{eqnarray}
where $\beta=(1-4m_f^2/M_{\Phi}^2)^{1/2}$ and  $p$ = 3(1) for the ${\cal
CP}$--even (odd) Higgs boson; the couplings $g_{\Phi ff}$ are listed in Tab.1.
For final state quarks one has to include QCD corrections; in the case of
$\CP$--even neutral Higgs bosons the correction factor is given by eq.~(2.5)
while in the case of the  pseudoscalar boson one has\cite{V4A,DG}
\beq
\Delta_{A}^{QCD} = \frac{1}{\beta}A(\beta) + \frac{1}{16\beta}(19+2\beta^2+
3 \beta^4)\log \frac{1+\beta}{1-\beta} +\frac{3}{8}(7 -\beta^2)
\eeq
with $A(\beta)$ given by eq.~(2.6). Here, again one has to use the running
masses which take into account the bulk of these QCD corrections, which in the
limit $M_H \gg m_q$ are the same for $\CP$--odd and $\CP$--even Higgs bosons;
as discussed previously in the case of $b$ quarks and for $M_{\Phi} \sim $ 100
GeV, this results in a decrease of the decay width by roughly a factor of two.
For values of $\tb$ larger than unity and for masses less than $\sim$ 130 GeV,
the	main decay modes of the neutral Higgs bosons will be decays into $b
\bar{b}$ and $\tau^+ \tau^-$	pairs;	the branching ratios being always
larger than $ \sim 90\%$ and $5\%$, respectively. The decays into $c \bar{c}$
and gluons are in general strongly suppressed especially for large values of
$\tb$. For large masses, the top decay channels $H,A \rightarrow t\bar{t}$ open
up; yet this mode remains suppressed for large $\tb$.

If the mass is high enough, the heavy ${\cal CP}$--even Higgs boson can in
principle decay into weak gauge bosons $H \rightarrow VV$, $V = W$ or $Z$.
Below the threshold for two real bosons, the ${\cal CP}$--even neutral Higgs
bosons $H$ and $h$ can decay into $VV^*$ pairs, one of the vector bosons being
virtual. The partial decay widths are given by
\begin{eqnarray}
\Gamma (\Phi \ra VV^{(*)}) =  g_{\Phi VV}^2 \Gamma (H_{\rm SM} \ra VV^{(*)})
\end{eqnarray}
\nn with $\Phi=h$~or~$H$ and $\Gamma (H_{\rm SM} \ra VV^{(*)})$ given by
eqs.~(2.9--2.11). Since $h$ is light and the $H$ partial width is proportional
to $\cos^2 (\beta-\alpha)$, they are strongly suppressed and the width of the
$ZZ$ signal is very small in the supersymmetric extension. [If $M_H$ is large
enough for these decay modes to be kinematically allowed, $M_h$ is very close
to its maximum so that $\cos^2 (\beta -\alpha) \rightarrow 0$.] For the same
reason, the cascade decay of the ${\cal CP}$--odd Higgs boson, $ A
\rightarrow	Zh$, is suppressed in general\cite{S2}
\begin{eqnarray}
\Gamma(A \ra Zh) = \frac{G_F}{8\sqrt{2} \pi} \cos^2 (\beta-\alpha)
\ \frac{M_Z^4}{M_A} \lambda^{1/2}(M_Z^2,M_h^2;M_A^2) \lambda(M_A^2,M_h^2;M_Z^2)
\end{eqnarray}
\nn with $\lambda(x,y;z)=(1-x/z-y/z)^2-4xy/z^2$ being the usual two--body phase
space function.

The heavy neutral Higgs boson $H$ can also decay into two lighter Higgs
bosons,\cite{S2}
\begin{small}
\begin{eqnarray}
\Gamma(H \ra hh) &=& \frac{G_F}{16\sqrt{2} \pi} \frac{M_Z^4}{M_H}
\left(1-4\frac{M_h^2}{M_H^2} \right)^{1/2} \left[ \cos 2\alpha \cos(\beta+
\alpha)-2 \sin 2\alpha \sin (\beta+\alpha)\right]^2 \nonumber \\
\Gamma(H \ra AA) &=& \frac{G_F}{16\sqrt{2} \pi} \frac{M_Z^4}{M_H}
\left(1-4\frac{M_A^2}{M_H^2} \right)^{1/2} \left[ \cos 2\beta \cos(\beta+
\alpha) \right]^2
\end{eqnarray}
\end{small}
\nn These modes, however, are restricted to small domains in
the parameter space.

Gluonic Higgs decays $\Phi \ra gg$ are mediated by top and bottom quark
loops [the squarks decouple from the effective $\Phi gg$ vertex for high
masses]. For the light Higgs particle this decay mode is significant only for
$h$ masses close to the maximal value where $h$ has $\SM$ like couplings, and
for $H$ masses only below 140 GeV and small values of $\tb$ where the coupling
to top quarks is sufficiently large. Therefore, one can neglect the $b$ loop
contribution, and the decay width $\Gamma(\Phi \ra gg)$ with $\Phi=h$ or $H$
is given by eq.~(2.12) up to the factor $g_{\Phi tt}^2$. For the pseudoscalar
Higgs particle, the gluonic decay mode is marginal.

Decays into of the $\CP$--even and $\CP$--odd Higgs bosons into $\gamma
\gamma$ and $Z \gamma$ final states are very rare with branching ratios of
order ${\cal O}(10^{-3})$ or below. The QCD corrections to the two photon decay
widths have been calculated in Ref.\cite{X4} and found to be small, $\sim
{\cal O}( \alpha_s/ \pi)$ across the physically interesting mass ranges if the
running of the quark masses is properly taken into account; for details see
section 4.3 where Higgs production in $\ga$ fusion will be discussed.

The	coupling of the charged Higgs particle to fermions is a
${\cal P}$ violating mixture of scalar and pseudoscalar couplings
\begin{eqnarray}
g_{H^+u\bar{d} } = \left( \frac{G_F}{\sqrt{2}}\right)^{1/2} \left[(1-\gamma_5)
 \frac{m_u}{\tb} + (1+\gamma_5) m_d \tb \right]
\end{eqnarray}

The charged Higgs particles decay into fermions with a partial decay width
\beq
\Gamma(H^+ \rightarrow u \bar{d}) &=&  N_c \frac{ G_F}{4 \sqrt{2} \pi}
\frac{\lambda^{
1/2}(m_u^2,m_d^2, M_{H^{\pm}}^2)}{M_{H^\pm}} \left[(M_{H^\pm}^{2} -m_{u}^{2}-
m_{d}^{2}) \right. \non \\
& & \left. \left( m_{d}^{2} {\rm tg}^2 \beta + \frac{m_u^2} {{\rm tg}^2 \beta}
\right) -4m_u^2m_d^2 \right]
\eeq
Here again one has to include the QCD corrections;\cite{DG,QCDHP} in the
limit $m_d=0$, corresponding to the approximate width in the case of the
top--bottom decays where the $m_b$ effects can be neglected, the QCD factor
is given by\cite{DG} [$\alpha=m_t^2/M_{H^\pm}^2$]
\beq
\Delta_{H^\pm}^{QCD} &=&  \frac{9}{4} + \frac{3-7\alpha+2 \alpha^2}
{2(1-\alpha)} \log \frac{\alpha}{1-\alpha} - \left[ 2{\rm Li }_2 \left( \alpha
\over \alpha -1 \right) \right. \non \\
& & \left. + \frac{\log (1-\alpha)}{1-\alpha} - \log (1-\alpha) \log{\alpha
\over 1-\alpha} \right]
\eeq

If allowed kinematically, charged Higgs bosons also decay into the lightest
neutral Higgs plus a $W$ boson,
\begin{small}
\begin{eqnarray}
\Gamma(H^{+} \rightarrow Wh) = 	\frac{G_F \cos^2 (\beta-\alpha) }{8\sqrt{2}\pi
c_W^2} \frac{M_W^4}{M_{H^\pm}} \lambda^{\frac{1}{2}}(M_W^2,M_H^2;M^2_{H^{\pm}})
\lambda(M_{H^\pm}^2, M_h^2; M_W^2) \hspace*{0.2cm}
\end{eqnarray}
\end{small}
Below the $tb$ and $Wh$ thresholds, the charged Higgs particles will decay
mostly into $\tau \nu_\tau$ and $c\bar{s}$ pairs, the former being dominant for
$\tb >1$.  For large $M_{H^\pm}$ values, the top--bottom decay $H^+
\rightarrow t\bar{b}$ becomes dominant.

Adding up the various decay modes,\cite{X5,PR3} the width of all five
Higgs bosons remains very small, even for large masses. This is shown for the
two representative values $\tb=2.5$ and 20 in Fig.~5. Apart from the ${\cal
CP}$--even heavy neutral Higgs boson $H$ and small $\tb$, the pattern of
branching ratios is in general quite simple. The neutral Higgs bosons decay
preferentially to $b \bar{b}$, and to a lesser extent to $\tau^+ \tau^-$ pairs;
the charged Higgs bosons to $\tau \nu_\tau$ and, preferentially, $t \bar{b}$
pairs above this threshold; see Fig.~6.

\vspace*{13.5cm}

\nn {\small Fig.~5 Total decay widths of the $\SUSY$ Higgs bosons [without
decays into $\SUSY$ particles] as functions of their masses for (a) ${\rm tg}
\beta=2.5$ and (b) ${\rm tg}\beta =20$. The top mass was chosen as $m_t=175$
GeV and $M_S=1$ TeV.}

\newpage

\vspace*{18cm}

\nn {\small Fig.~6 Decay branching ratios of the Higgs bosons [without $\SUSY$
decays] as functions of their masses for two values of ${\rm tg}\beta=2.5$ and
20; $m_t=175$ GeV and $M_S=1$ TeV.}

\newpage

\vspace*{6cm}

\nn \begin{center} {\small Fig.~6 (continued).} \end{center}

\vspace*{0.6cm}

In most studies of supersymmetric Higgs bosons at future colliders it is
assumed that they do not decay into supersymmetric particles. However, while
sfermions are probably too heavy to affect Higgs decays, the Higgs boson
decays into charginos and neutralinos could eventually play a significant role
since some of these particles are expected to be lighter or of the same order
as $M_Z$. These new channels could open up at least for the heavy Higgs bosons
$H,A$ and $H^\pm$.\cite{V12} They could be so large that they reduce the
branching fractions for the standard decays in a sizable way, therefore
altering the signals, and as a result, the search  strategies for these
particles.\cite{X5,Baer}

The decay widths of the neutral Higgs bosons $H_k$, with $H_1=H$, $H_2=h$
and $H_3=A$, into chargino or neutralino pairs are given by\cite{V12}
\begin{small}
\begin{eqnarray}
\Gamma(H_k \ra \tilde{\chi}_i \tilde{\chi}_j) = K \frac{M_H}{\delta(i,j)}
\left[(F_{ijk}^2+F_{jik}^2)
\left(1-\frac{M_i^2} {M_H^2}-\frac{M_j^2}{M_H^2}\right)-4F_{ijk}F_{jik}\eta_k
\frac{|M_iM_j|}{M_H^2} \right]
\end{eqnarray}
\end{small}
\hspace*{-2mm}where $\eta_{1,2}=+1, \ \eta_3=-1$ and $\delta(i,j)=1$
unless the final state consists of two identical Majorana neutralinos in which
case $\delta(i,j)=2$; $K=G_F/(2\sqrt{2} \pi)\lambda^{1/2} M_W^2$.
The coefficients $F_{ijk}$ can be expressed in terms of
the elements of the matrices $U,V$ which diagonalize the chargino mass
matrices, and of the neutralino matrix $Z$, given in Ref.\cite{V12} For the
charged Higgs boson decays into neutralino/chargino pairs, the partial
widths read [$F_{L,R}$ can be found in Ref.\cite{V12}]
\begin{small}
\begin{eqnarray}
\Gamma(H^\pm \ra \tilde{\chi}_i^\pm \tilde{\chi}_j^0)= K
M_{H^\pm} \left[ (F_{L}^2+F_{R}^2)\left(1-\frac{M_i^2}
{M_{H^\pm}^2}-\frac{M_j^2}{M_{H^\pm}^2} \right)-4F_{L}F_{R} \frac{|M_iM_j|}
{M_{H^\pm}^2}  \right]
\end{eqnarray}
\end{small}

The decay branching ratios of the lightest ${\cal CP}$--even neutral Higgs
boson $h$ into the lightest chargino pair and the lightest and
next--to--lightest neutralino pairs, are displayed in Fig.~7, an
update of Ref.\cite{X5,inv} The
contours are shown in the [$\mu,M$] plane where the sum of these branching
ratios exceeds 5\% [dashed lines] and 50\% [solid lines]; the dotted lines are
the countours which are excluded by LEP100 data and which can be probed at
LEP200. These decays can be sizable in the area of the $[\mu,M]$ plane between
the two LEP contours. They are particularly important for $h$ masses close to
the maximum allowed values [see Fig.~2]; in this case the lightest Higgs boson
has $\SM$ couplings and the dominant $b\bar{b}$ decay mode is not enhanced
anymore for large $\tb$ values, so that other decay modes can become
significant. For these masses and for large $\tb$ values, the branching ratios
for neutralino/chargino decays are sizable even outside the regions which can
be probed at LEP200.

The branching ratios of the heavy ${\cal CP}$--even, the ${\cal CP}$--odd
and the charged Higgs boson decays into chargino and neutralino pairs can be
very large and they can even be dominant in some areas of the $\MSSM$ parameter
space, exceeding 50\% for positive values of $\mu$ and/or $M$ values below 200
GeV.\cite{X5} This is mainly due to the fact that the couplings of the Higgs
bosons to charginos and neutralinos are gauge couplings which can be larger
than the Yukawa couplings to standard fermions and the couplings to the gauge
bosons [the latter being zero at tree level for the ${\cal CP}$--odd and being
suppressed for the heavy ${\cal CP}$--even Higgs particles]. Finally, the
branching fractions of the invisible neutral Higgs decays can be important,
Fig.~8,  and they could jeopardize the search for the Higgs particles at hadron
colliders.

\vspace*{9.5cm}

\nn {\small Fig.~7. Contour lines in the $[\mu,M$] plane where the sum of the
branching ratios of the lightest Higgs boson $h$ into charginos and neutralinos
exceeds 5\% (dashed lines) and 50\% (full lines) for two values of $M_h$
and $\tb$; the shaded areas are the regions which are (can be) excluded at
LEP100 (LEP200). The top mass was taken to be $m_t=140$ GeV.}

\newpage

\vspace*{18cm}

\nn {\small Fig.~8. Branching ratios of the decays of the three neutral Higgs
bosons into the lightest neutralino pair [invisible decays] as a function of
their masses for $\tb=2.5$ and 20. The top mass is fixed to 175 GeV
and $M_S=1$ TeV.}

\renewcommand{\theequation}{3.\arabic{equation}}
\setcounter{equation}{0}

\section{Production at pp Colliders}

\vspace*{-1mm}

\subsection{Standard Model Higgs}

\nn 3.1.1. {\it Production mechanisms} \s

\nn The major goal of the CERN future ``Large Hadron Collider" [and the one of
the late ``Superconducting Super Collider" planned in Texas] is the search for
Higgs particles. There are four main production mechanisms of neutral Higgs
bosons at hadron colliders, and all of them make use of the fact that the Higgs
boson couples preferentially to heavy particles. These four processes, see
Fig.~9, are the following:
\beq
{\rm gluon-gluon~fusion~mechanism}: & & gg  \ra H \hspace*{2cm} \\
{\rm massive~vector~boson~fusion}: & & WW/ZZ \ra   H \\
{\rm associate~production~with}~W/Z: & & q\bar{q} \ra V + H \\
{\rm associate~production~with}~\bar{t}t: & & gg,q\bar{q}\ra t\bar{t}+H
\eeq

\vspace*{7cm}

\nn \centerline{\small Fig.~9 Main production mechanisms of Higgs bosons at
proton colliders.\s}}

The cross sections\cite{P4}  are shown in Fig.~10 for a center of mass energy
of 16~TeV, typical of LHC. [Note that for a luminosity of ${\cal L}=10^{33}
(10^{34})$~cm$^{-2}$s$^{-1}$, $\sigma=$~1 pb would correspond to $10^{4}
(10^{5})$ events per year.]

All the way up to Higgs masses of the order of 1 TeV, the dominant production
process is the gluon--gluon fusion mechanism\cite{V11} which proceeds through a
triangular top quark loop in the Standard Model. In the intermediate Higgs mass
range $M_W<M_H<2M_Z$, the cross section is of the order of a few tens of pb [in
Fig. 10 the two extreme values of the top mass, $m_t=150$ and $m_t=200$ GeV,
are used]. The cross section is enhanced for Higgs masses near the $2m_t$
threshold and drops down to values of the order of $0.1$ pb for $M_H\sim$ 1
TeV.

The next most important production mechanism is the $WW/ZZ$ fusion
process,\cite{P1} $q q \ra V^* V^* \ra qq H$. For intermediate mass Higgs
bosons, the cross section is of the order of a few pb, i.e. one order of
magnitude below the $gg$ fusion mechanism cross section, but it drops less
rapidly than the $gg \ra H$ cross section and becomes competitive with the
latter for Higgs masses of the order of 1 TeV. In fact, when the top quark was
believed to be rather light, this process has received much attention since it
was the dominant production mechanism in a wide Higgs mass range.

\vspace*{10cm}

\nn {\small Fig.~10  Cross sections [in pb] for $\SM$ Higgs production at
the LHC as a function of the Higgs mass. The c.m. energy was taken to be
$\sqrt{s}=16$ TeV; for the processes $gg \ra H$ and $pp \ra t\bar{t}H$, the top
quark mass was chosen to be $m_t=150$ and 200 GeV.} \s

The process where the Higgs particle is produced in association with $W/Z$
gauge bosons\cite{P2} is important only for intermediate mass Higgs bosons
where
the cross sections are another order of magnitude smaller than the one for
$WW/ZZ$ fusion. However, if the luminosity is high enough, this process can be
useful since one can trigger on the $W$ or $Z$ bosons through their leptonic
decay modes. Note that the $WH$ process dominates over the $ZH$ process, a
consequence of the fact that the charged current couplings are larger than the
neutral couplings.

Finally, the Higgs production in association with $t\bar{ t}$ pairs\cite{P3}
has a cross section of the same order as the $q \bar{q} \ra VH$ cross section
in the intermediate mass range, i.e. ${\cal O}$ (1~pb) [for $m_t=175$ GeV, the
cross section would be slightly smaller than shown in the figure]. The process
proceeds through gluon--gluon fusion as well as $q\bar{q}$ annihilation, the
contribution of the former being much larger than the one of the latter. As for
the $q \bar{q} \ra VH$ process, it could be useful in the intermediate mass
range since one can also trigger on the $W$ boson produced in the $t \ra b W$
decay.

Note that for the SSC with a planned c.m. energy of 40 TeV, the cross
sections\cite{P5} would have been approximately 3 times larger than at the LHC.

For completeness, we will give expressions for the matrix element squared,
once the
sum/average over spin/color is performed, for the two processes $q \bar{q} \ra
VH$ and $qq \ra V^* V^* \ra qqH$ where $V=W,Z$ [note that there are a few
misprints in the corresponding formulae in the compilation of the first paper
in Ref.~\cite{LHC}].

For the $q \bar{q} \ra VH$ process, one has
\beq
|{\cal M}(q \bar{q} \ra VH)|^2 = \frac{G_F^2 M_V^4}{18 \hat{s}} \ (v_q^2
+a_q^2)
\frac{\lambda(M_V^2/\hat{s}, M_H^2/\hat{s} ) +12M_Z^2/\hat{s} }{(1-M_V^2/
\hat{s})^2}
\eeq
where $\lambda$ is the usual two--body phase space function $\lambda(x,y)=
(1-x-
y)^2-4xy$ and $v_q/a_q$ the vector/axial--vector couplings of the quarks to the
vector bosons: $v_q=\sqrt{2}, a_q=-\sqrt{2}$ for the $W$ and $v_q=2I_{3q}
-4e_qs_W^2$ and $a_q=2I_{3q}$ for the $Z$ [$I_{3q}$ is the third component of
weak isospin and $e_q$ the electric charge of the quark].

In the case of the $qq \ra qqH$ production mechanism, one has in terms of the
scalar products of the quark momenta
\beq
|{\cal M}|^2 \left( q (p_1) q(p_2) \ra q(p_1^,) q(p_2^,) H \right)
=  \frac{32 \sqrt{2} G_F^3 M_V^8}{[(p_1-p_1^,)^2-M_V^2]^2[(p_2-p_2^,)^2-
M_V^2]^2} \non \\
\times \left[~ (g_L^2 g_L^{,2} + g_R^2 g_R^{,2}) (p_1 \cdot p_2 \ p_1^{,}
\cdot p_2^{,} ) +(g_L^2 g_R^{,2} + g_R^2 g_L^{,2}) (p_1 \cdot p_2^, \ p_1^{,}
\cdot p_2) ~\right]
\eeq
where $g_{L,R}= (v_q \mp a_q)/2$ with $v_q$ and $a_q$ defined as previously.
To obtain the total cross--sections, one has to sum over the contributions of
quarks and antiquarks [with the appropriate couplings] and convolute with the
appropriate quark distribution functions. The cross section for the $pp \ra
t\bar{t}H$ is rather involved, the $gg$ fusion mechanism will be discussed
separately in the next subsection.

Besides the errors due to the poor knowledge of the gluon distribution at small
$x$ [this will be improved by future measurements at the $ep$ collider HERA],
the lowest order cross sections are affected by large uncertainties due to
higher order corrections. Including the next to leading QCD corrections, the
total cross sections can be defined properly: the scale at which one defines
the strong coupling constant is fixed and the [generally non--negligible]
corrections are taken into account. The ``K--factors" for $VH$
production\cite{P6} [which can be inferred from the Drell--Yan production of
weak bosons] and the $VV$ fusion mechanisms\cite{P6P} are small, increasing the
total cross sections by $\sim$ 20\% and 10\% respectively;  the corrections to
the associate $t\bar{t}H$ production process are still not known. The QCD
radiative corrections to the main production mechanism, $gg\ra H$, have been
computed in Refs.\cite{X3,P7} and have been found to be rather large; they will
be discussed in some detail in the following.

\nn 3.1.2. {\it gg fusion mechanism} \s

\nn As previously mentioned, the dominant production mechanism for the Standard
Higgs boson at proton--proton colliders is the gluon--gluon fusion
process.\cite{V11} The Higgs boson couple to gluons primarily through a heavy
top quark triangle loop. To lowest order ($LO$), the cross section is given by
\begin{equation}
   \sigma_{LO}(pp\rightarrow H +X)=\sigma_0^H
       \, \tau_H \frac{d{\cal L}^{gg}}{d\tau_H}
\end{equation}
where $d {\cal L}^{gg} / d \tau_H$ denotes the gluon luminosity
at $\tau_H = M^2_H / s$ with $\sqrt{s}$ being the c.m. energy of the
proton collider. The parton cross sections can be expressed in terms
of a form factor derived from the quark triangle diagrams in Fig.~9,
\begin{eqnarray}
\sigma_H & = & \frac{G_F \alpha_s^2}{288 \sqrt{2}\pi}
               \left|\sum_Q F_Q^H (\tau_Q) \right|^2
\end{eqnarray}
The form factor is related to the scalar triangle integral $f$,
\begin{eqnarray}
f(\tau_Q) & = & \left\{
               \begin{array}{ll}
               \arcsin^2\sqrt{\tau_Q} & \tau_Q<1\\
               -\frac{1}{4}\left[
                              \log\frac{1+\sqrt{1-\tau_Q^{-1}}}
                                       {1-\sqrt{1-\tau_Q^{-1}}}-i\pi
                           \right]^2 & \tau_Q>1
               \end{array}
             \right.
\end{eqnarray}
in the following way
\begin{eqnarray}
F_Q^{H} (\tau_Q) & = & \frac{3}{2}\tau_Q^{-1} \left[
                            1+(1-\tau_Q^{-1})f(\tau_Q) \right]
\end{eqnarray}
with $\tau_Q = M^2_H / 4 m^2_Q$. The form factor is normalized such that for
$m_Q \gg M_H$, $F^H_Q \ra 1$ and it approaches zero in the chiral limit $m_Q
\ra 0$.
To incorporate the QCD corrections to $\sigma (pp \ra H + X)$, one has to
consider the processes,
\begin{equation}
gg \ra H (g)  \hspace{5mm} \mbox{and} \hspace{5mm}
gq \ra H q,   \hspace{5mm} q \overline{q} \ra H g
\end{equation}
Characteristic diagrams of the QCD radiative corrections are shown
in Fig.~11. They consist of two--loop gluon--quark and
Higgs--quark vertex corrections, rescattering corrections and
non--planar diagrams. The renormalization program has been carried
out in the $\overline{\rm MS}$ scheme. The ``physical" quark mass
$m_Q$ is defined at the pole of the propagator;
this assures the correct [perturbative] threshold behavior of the
triangle amplitude for $M_H \approx 2 m_Q$. The
renormalization of the scalar $HQ\overline{Q}$
vertex is connected with the renormalization of the quark mass and the
quark wave--function, $Z_{HQQ}  = (Z_Q -1) - \delta m_Q/m_Q$.\cite{V4}
In addition to these virtual
corrections, the gluon radiation off the initial state gluons
and the heavy quark lines must be taken into account. After
adding up all these contributions, ultraviolet and infrared
singularities cancel. Leftover collinear singularities
are absorbed into the re\-nor\-ma\-lized parton densities\cite{N9}
which we define in the $\overline{\rm MS}$ scheme. Finally the
subprocesses $gq \ra Hq$ and $q \overline{q} \ra Hg$ must be
added.

\newpage

\vspace*{8cm}

\nn {\small Fig.~11 Generic Feynman diagrams for the QCD corrections to Higgs
boson production in the $gg$ fusion mechanism.} \\

The results for the cross sections can be summarized in the following
form:
\begin{equation}
\sigma(pp\rightarrow H +X)=\sigma_0^H
         \left[
            1+C^H \frac{\alpha_s}{\pi}
         \right] \tau_H
         \frac{d{\cal L}^{gg}}{d\tau_H}
         +\triangle \sigma_{gg}^H
         +\triangle \sigma_{gq}^H
         +\triangle \sigma_{q \overline{q}}^H
\end{equation}
The coefficient $C^H$ denotes the contributions from the virtual
two--loop quark corrections regularized by the infrared singular
part of the cross section for real gluon emission,
\begin{eqnarray}
C^H &=& \pi^2 + c^H + \frac{33-2 N_f}{6} \log
\frac{\mu^2}{M^2_H } \nonumber \\
c^H  &=& Re \sum_Q F^H_Q \,c^H_Q (\tau_Q) / \sum_Q
F^H_Q
\end{eqnarray}
The coefficient $C^H$
splits into the infrared term $\pi^2$, a term depending on the
re\-nor\-ma\-li\-za\-tion scale $\mu$
of the coupling constant, and a piece $c^H$ which depends on the mass ratio
$\tau_Q$; it has been reduced from
5--dimensional Feynman parameter integrals to 1--dimensional
integrals analytically and the remaining integration has been
performed numerically. In the limit of large quark masses, the coefficient
can be calculated analytically and one finds,\cite{X3,P7}
\begin{eqnarray}
\mbox{$m_Q \gg M_H$:} \hspace{2cm} c^H_Q \ra 11/2
\end{eqnarray}
For large Higgs masses but moderate quark
masses the real and imaginary parts of $c^H_Q$ remain small,
$\leq \pm 5$ for $\tau_Q  \leq 10^4$. However, the numerical results
clearly indicate the onset of the asymptotic behavior
${\cal I}m c^H_Q \sim \log \tau_Q$
and suggest ${\cal R}e c^H_Q \sim \log^2 \tau_Q$ for large
$\log \tau_Q$.
The (non--singular) contributions from gluon radiation in $gg$
scattering, from $gq$ scattering and $q \overline{q}$ annihilation,
Figs.~11, depend on the renormalization scale $\mu$ and the factorization
scale $M$ of the parton densities,
\begin{eqnarray}
\triangle \sigma_{gg}^H & = &
       \int_{\tau_H}^1 d\tau \frac{d{\cal L}^{gg}}{d\tau}
       \frac{\alpha_s}{\pi} \sigma_0^H
       \left\{-z P_{gg}(z)\log\frac{M^2}{\tau s}
       +d_{gg}^H (z,\tau_Q) \right.  \nonumber \\
& & \hspace{2.2cm} \left.
                  +12 \left[\left(\frac{\log(1-z)}{1-z}\right)_{+}
                  -z\left[2-z(1-z)\right]\log(1-z) \right]
       \right\} \nonumber \\
   \triangle\sigma_{gq}^H & = &
       \int_{\tau_H}^1 d\tau \sum_{q,\overline{q}}
       \frac{d{\cal L}^{gq}}{d\tau}
       \frac{\alpha_s}{\pi} \sigma_0^H
       \left\{ \left[ -\frac{1}{2}\log\frac{M^2}{\tau s}+\log(1-z)
       \right] z P_{gq}(z) +d_{gq}^H (z,\tau_Q) \right\} \nonumber\\
   \triangle\sigma_{q\overline{q}}^H & = &
       \int_{\tau_H}^1 d\tau \sum_q
           \frac{d{\cal L}^{q\overline{q}}}{d\tau}
       \frac{\alpha_s}{\pi} \sigma_0^H
       d_{q\overline{q}}^H (z,\tau_Q)
\end{eqnarray}
with $z=\tau_H/\tau$ and $P_{ij}$ being the standard Altarelli--Parisi
splitting functions.\cite{AP} Again, the coefficients $d^H_{gg}, d^H_{gq}$ and
$d^H_{q \overline{q}}$, have been reduced to 1--dimensional integrals which
have been evaluated numerically. In the limit of large
quark masses, the coefficients can be determined analytically\cite{X3,P7}
\begin{eqnarray}
\mbox{$m_Q \gg M_H$:} \hspace{2cm}
d_{gg}^H & \rightarrow & -\frac{11}{2}(1-z)^3
\hspace*{4cm} \nonumber \\
d_{gq}^H & \rightarrow & -1+2z-\frac{1}{3}z^2 \nonumber \\
d_{q\overline{q}}^H & \rightarrow & \frac{32}{27}(1-z)^3
\end{eqnarray}

In Fig.~12 we present the final QCD corrected cross sections for a value of
the top mass of 200 GeV, assuming that the Higgs mass is sufficiently below the
$2m_t$ threshold. The cross sections are shown for the DFLM
parametrization\cite{DFLM} of the parton densities and of the coupling
constant in the
next--to--leading order at $\mu^2=M^2=\hat{s}$ and also $\mu^2=M^2=M_H^2$
with $\Lambda^5_{\overline{MS}}= 170$ MeV. They are compared with the Born
terms in which all quantities are consistently used in the leading order
expansion $\Lambda_{\rm LO}^5=130$ MeV. The QCD corrections are found to be
at the level of 50 to 80\%.  To demonstrate the uncertainty of the cross
section
due to the presently unknown small $x$ behavior of the gluon density and the
magnitude of the coupling constant, we compare the predictions of the
DFLM\cite{DFLM} and GRV\cite{GRV} parametrizations; even for these two
extremes choices, the difference remains at a level less than 20\%.

The QCD corrections to $gg \ra H$ for arbitrary values of $M_H^2/m_t^2$
have been calculated in Refs.\cite{P10} and the result is shown in Fig.~13.
As one can see, the $K$ factor which characterizes the size of the QCD
corrections, $K^H=\sigma_{\rm HO}^H / \sigma_{\rm LO}^H$, is rather insensitive
to the mass ratio and is of order 1.8. Therefore the result in the limit $m_t
\gg M_H$ is a good approximation of the complete result.

\newpage

\vspace*{7.5cm}
\nn {\small Fig.~12 Production cross sections with and without the QCD
corrections at the LHC in the approximation $m_t \gg M_H/2$; the top mass
is fixed to 200~GeV.} \\

\vspace*{8.7cm}

\nn {\small Fig.~13 Individual and total contributions to the ``K--factor"
of the $gg \ra H$ process at the LHC as a function of the Higgs mass; the top
mass is fixed to 175~GeV. The GRV parametrization for the parton densities has
been used.} \\

The next important radiative correction to the $Hgg$ amplitude, that is
proportional to the square of the mass of the quark in the loop and is
therefore
potentially very large, is the two--loop ${\cal O}(G_F m_Q^2)$ electroweak
correction. We will summarize below the result of a very recent evaluation of
this correction.\cite{ggEW}

As already discussed, in the minimal Standard Model with three families, only
the top quark significantly contributes to the $Hgg$ coupling. The ${\cal
O}(\alpha_S G_F m_t^2)$ correction to the top quark loop amplitude in the limit
$m_t \gg m_b$ is given by\cite{ggEW}
\begin{eqnarray}
F_t^H \ra F_t^H \left[1 + \frac{G_F \sqrt{2}}{32 \pi^2} m_t^2 \right]
\end{eqnarray}
Due to a large cancellation among the various contributions, the total
correction amounts to a positive contribution of a mere 0.2\% for a top mass
value $m_t \sim 200$ GeV. Therefore, contrary to the very large QCD
corrections, the leading electroweak correction to the top quark loop mediated
Higgs--gluons coupling turns out to be very small and well under control.

Since the measurement of the $Hgg$ coupling
is a very powerful tool to count the number
of heavy  quarks which couple to the Higgs boson, it is interesting to evaluate
the effect of this correction in the case of a fourth generation of heavy
fermions. Because the mass splitting between the members of the extra weak
isodoublet of quarks is highly constrained by present electroweak precision
measurements, one can work in the approximation where the two quarks are
degenerate in mass. In this case, the ${\cal O}(G_F m_Q^2)$ correction to one
of the quarks amplitude is given by\cite{ggEW}
\begin{eqnarray}
F_Q^H \ra F_Q^H \left[1 - \frac{G_F \sqrt{2}}{ 8\pi^2} m_Q^2 \right]
\end{eqnarray}

This negative correction will therefore screen the value of the one--loop
generated $Hgg$ coupling. However, the correction is rather small since for
realistic values of the quark masses, $m_Q < 500$ GeV [an upper bound obtained
from the requirement that weak interactions do not become strong and
perturbation theory is reliable\cite{CFH}], it does not exceed the 5\% level.
It is only for quark masses larger than $\sim 2$ TeV, for which perturbation
theory breaks down already at the tree level, that this radiative correction
will exceed the one--loop result.

Note that in the previous equation only the contribution of the heavy quarks of
the fourth generation has been taken into account. Additional contributions
will be induced by the extra weak isodoublet of leptons [from the
renormalization of the Higgs boson wave--function and vacuum expectation
value]. If one assumes that the masses of the heavy leptons are approximately
equal to those of the quarks, the total correction in eq.~(3.18) will be
smaller by a factor of three.

Therefore, the ${\cal O}(G_F m_Q^2)$ correction to the $Hgg$ amplitude is well
under control for quark masses in the range interesting for perturbation
theory, and the counting of new heavy quarks via the $Hgg$ coupling will not be
jeopardized by these corrections. \\

\nn 3.1.3. {\it Higgs Detection} \s

\nn The signals which are best suited to identify the Higgs boson at proton
colliders have been studied in great detail; see for instance
Refs.\cite{P4,P5}.
Here, we will briefly summarize the main conclusions of these studies for
LHC energies. For the sake of convenience we will divide the Higgs mass into
two ranges: the ``high mass" range $M_H > 140$ GeV and the ``low mass" range
$M_H <140$ GeV. \\

\vspace*{-2mm}

\nn \underline{High mass range:} \\

\vspace*{-2mm}

\nn For a Higgs mass above the $ZZ$ threshold, $M_H > 2M_Z$, one can exploit
the $H\ra ZZ$ decays which have clean signatures and are affected by rather
small backgrounds.  The best signal is the so called ``gold--plated"
signal\cite{GKW} $H \ra ZZ \ra 4 l^\pm$ with two pairs of charged electrons or
muons with invariant masses equal to $M_Z$. The branching ratio is of the order
of $\sim 0.14\%$ and is therefore somewhat small. To improve the signal cross
sections, one can also use the $H \ra l^+l^- \bar{\nu} \nu$ decay mode for
which the branching ratio is 6 times larger; however in this latter mode the
Higgs mass cannot be measured directly.

The backgrounds, which come mainly from $Z$ boson pair production in $q \bar{q}
\ra ZZ$ and in the loop mediated process $gg \ra ZZ$ [for the $H \ra l^+l^-
\bar{\nu} \nu$ mode one has to consider in addition the process $pp \ra Z+$
jets, where some of the jets are undetected or their energy not well measured
to give a fake missing energy] are manageable: they have softer transverse
momentum distributions than to the $H \ra ZZ$ signal for which $P_T(Z) \sim
M_H/2$. For a total integrated luminosity of $10^5$ pb$^{-1}$, one can detect
the Standard Model Higgs boson up to masses of $M_H \sim 800$ GeV with the
$4l^\pm$ signal, but for a luminosity 10 times smaller only masses $M_H \sim
500$ GeV can be reached. The complementary signal $H\ra l^+l^- \bar{\nu}\nu$ as
well as the process $H\ra WW\ra l\nu jj$, where the Higgs boson is produced in
the $VV$ fusion processes and where one uses forward jet tagging, can extend
the discovery limit up to Higgs masses of 1 TeV [only $M_H \sim 700$ GeV for an
integrated luminosity of $10^4$ pb$^{-1}$].

Below the $2M_Z$ threshold, $H \ra ZZ^* \ra 4l^\pm$, where one of the
$Z$ bosons is off--shell, has a still appreciable cross section times branching
ratio. The signal is very clean and after appropriate transverse momentum
cuts on the charged leptons and with a good lepton isolation and
identification,
one can suppress the backgrounds [which mainly come from $ZZ^*,Zb\bar{b},Zt
\bar{t}$ and $t\bar{t}$ production, the heavy quarks decaying
semi--leptonically] down to a manageable level. Higgs bosons with masses down
to $M_H \simeq 140$ GeV can be detected in this process.

One can therefore conclude that with the $H \ra ZZ^{(*)} \ra$ 4 leptons signal,
the $\SM$ Higgs boson with a mass in the ``high range", 140 GeV $ < M_H < 1$
TeV can be relatively easily detected at the LHC, provided that the
luminosity is high enough, i.e. ${\cal L}=10^{34}$~cm$^{-2}$s$^{-1}$. The high
luminosity is demanding on the detector, though. \\

\vspace*{-2mm}

\nn \underline{Low mass range:} \\

\vspace*{-2mm}

\nn For Higgs bosons in the ``low mass" range, the situation is a bit more
complicated. The branching ratio for $H\ra ZZ^*$ becomes too small to be useful
and because of the huge QCD jet background, the dominant Higgs decay mode $H
\ra b\bar{b}$ is useless in the main production process $gg \ra H$; one has
then to rely\cite{GKW} on the very rare $\gamma \gamma$ decay mode with a
branching ratio of ${\cal O}(10^{-3})$.

At the LHC with a luminosity of $\int {\cal L}= 100$~fb$^{-1}$, the cross
section times branching ratio leads to ${\cal O}(0.5-1 \times 10^{3})$ events
in the mass range $80 <M_H < 140$ GeV but one has to fight against formidable
backgrounds. Jets faking photons need a rejection factor larger than $10^{4}$
per jet to be reduced to the level of the physical background $q\bar{q}, gg \ra
\gamma \gamma$ which is still very large. However, if very good geometric
resolution and stringent isolation criteria, combined with excellent
electromagnetic energy resolution to detect the narrow $\gamma \gamma$ peak of
the Higgs boson are available [one also needs a high luminosity $ {\cal L}
\simeq 10^{34}$~cm$^{-2}$s$^{-1}$], this channel, although very difficult, is
feasible.

A complementary channel in this mass range would be the $q \bar{q} \ra WH,
t\bar{t}H\ra \gamma \gamma l \nu$ using the charged lepton from the $W$ decay
as a tag. At LHC, the signal for the associate production of the Higgs with
top quark pairs is two times larger than the one for associate production with
a $W$. The background cross sections are rather small but the signal cross
sections are also small: only $ \sim 20$ events for a luminosity of $10^{5}$
pb$^{-1}$ are expected [when one adds up the two channels], making this process
also difficult to use.

Two alternatives of detecting the Higgs boson through decays which are not
the two--photon decays have been proposed. The first one is the $H \ra \tau^+
\tau^-$ channel\cite{RKEllis} with the Higgs boson produced in the gluon--gluon
fusion mechanism. Unfortunately, because of overwhelming physical backgrounds
[coming mainly from $t\bar{t}$ pair production and also from the Drell--Yan
production of $\tau^+ \tau^-$ pairs for $M_H$ in the vicinity of $M_Z$], this
method has been found hopeless for the Standard Model Higgs.\cite{dilela}
However, it could be used in the supersymmetric extension for $H$ and $A$
in some range of the $\SUSY$ parameter space.

A second method which has been proposed very recently, is to use the dominant
$H \ra b\bar{b}$ decay mode with the Higgs particles produced in association
with $t\bar{t}$ pairs,\cite{P11} leading to $ttbb$ final states. Requiring that
at least one of the top quarks decays into an isolated $e$ or $\mu$ lepton and
with the help of a micro--vertex detector with very good efficiency and purity
for tagging the $b$ quarks, it would be possible to use this channel at the LHC
for a top quark heavier than 150 GeV. However, because of the many overlapping
events expected at the LHC with a very high--luminosity option [which is needed
to have a reasonable number of events], the efficiencies for $b$--quark tagging
might be difficult to achieve at the required level.\cite{P11} A careful
analysis including the simulation of the experimental environment\cite{X6}
might be required to assess firmly the viability of this mode.

In conclusion, while it is relatively easy to detect Higgs bosons in the
``high mass" range up to $\sim 1$ TeV at LHC, it is rather difficult to detect
Higgs bosons with masses below $140$ GeV: a very high luminosity $\simeq
10^{34}$~cm$^{-2}$s$^{-1}$ and a dedicated detector are required.

\newpage

\subsection{Supersymmetric Extension}

\nn 3.2.1. {\it Production mechanisms} \s

\nn In the minimal supersymmetric extension of the Standard Model, the
mechanisms for producing the neutral Higgs bosons are practically the same as
in the $\SM$, one simply has to take into account the contributions of the $b$
quark whose couplings are in general stronly enhanced for large $\tb$ values,
and for the pseudoscalar Higgs boson, discard the $WW/ZZ$ fusion processes and
the associate production process with a $W$ or $Z$ boson because there is no
$AVV$ coupling at the tree level. The main features of these production
mechanisms are summarized below:\cite{N1a,N1}

(a) $gg \ra h/H/A$: in the $\MSSM$ this process proceeds through top and
bottom quark loops. Because the top (bottom) couplings to the $\SUSY$ Higgs
bosons are suppressed (enhanced) compared to those of the $\SM$ Higgs, the $b$
quark loop contribution can compete with the one of the top quark and
becomes even
 dominant for large values of $\tb$. Extra contributions may come from
squark loops, but contrary to quarks, these particles will decouple from the
Higgs--gluon vertex if their masses are very large. The rather large QCD
corrections to this process will be discussed in the next subsection.

(b) $WW/ZZ$ fusion mechanisms: in the $\MSSM$, the couplings of the
$\CP$--even Higgs bosons to $VV$ pairs are suppressed by $\sin(\alpha-\beta)$
or $\cos( \alpha-\beta)$ factors compared to the $\SM$ couplings and the
production cross sections are therefore always smaller than in the $\SM$. In
fact, this process is not important in general since for large Higgs masses
where it dominates over the $gg$ fusion mechanism in the $\SM$, the $HVV$
coupling is extremely small; for small Higgs masses, the $hVV$ couplings can
approach the $\SM$ value but the cross section is much below the one for the
$gg$ fusion process. In the case of the pseudoscalar Higgs boson, because there
is no $AVV$ coupling at the tree level, these processes are absent.

(c) Associate production with $W/Z$: this process is important only for
the lightest Higgs boson $h$ and for the heavier $H$ with a mass in the ``low
mass" range [again, because there is no $AVV$ coupling this process is absent
for the pseudoscalar]. Because of the suppressed $h/HVV$ couplings, the cross
sections are always smaller than in the case of the $\SM$ Higgs.

(d) Associate production with a heavy quark pair: because the top quark
couplings to Higgs bosons are smaller than in the $\SM$ for $\tb >1$, the cross
section for the process $q\bar{q}, gg \ra Q\bar{Q} + h/H/A$ where $Q=t$ is
smaller than the $\SM$ Higgs cross section. However, for $Q=b$ this process
can become the dominant production mode in some range of the $\SUSY$ parameter
space; this is because the $b$ couplings to Higgs bosons are extremely enhanced
for large values of $\tb$.

(e) For charged Higgs bosons, the most interesting production processes are:
the top decay\cite{Htb} if $M_{H\pm} <m_t-m_b$, $t \ra H^+b$ with the top
quarks mainly produced through gluon fusion, $gg \ra \bar{t}t$ and
possibibly\cite{gbtH} $gb \ra tH \ra ttb$ in a limited range of the $\SUSY$
parameter space. Other production processes seem to be hopeless at hadron
colliders.

\nn 3.2.2. {\it gg fusion mechanism} \s

\nn In this subsection we discuss the gluon--gluon fusion production mode for
supersymmetric neutral Higgs bosons and present the QCD corrections to the
production rates.\cite{X7} This demands the extension of the $\SM$
calculation in two ways: (i) since the Higgs--$gg$ coupling is mediated, in
part of the $\MSSM$ parameter space, dominantly by bottom quark loops one has
to deal with a properly weighted superposition of $t$ and $b$ loop
contributions, (ii) the QCD corrections have to be analyzed for arbitrary
ratios of Higgs and quark masses, and (iii) the QCD corrections are to be
determined for pseudoscalar Higgs bosons.

To lowest order ($LO$), the cross sections for the production of Higgs
particles in $gg$ fusion are given by eq.~(3.7) where the Standard Model
Higgs boson $H$ has to be replaced by $\Phi=h/H$ or $A$. For the $\CP$--even
Higgs bosons $h$ and $H$ the parton cross section is given by eq.~(3.8), while
for the pseudoscalar one has
\begin{eqnarray}
\sigma_0^A        & = & \frac{G_F \alpha_s^2}{128 \sqrt{2}\pi}
               \left|\sum_Q F_Q^A(\tau_Q) \right|^2
\end{eqnarray}
With the help of the scalar triangle integral $f$ the form factors $F_Q^{h/H}$
and $F_Q^A$ are
\begin{eqnarray}
F_Q^{h/H} (\tau_Q) & = & \frac{3}{2}\tau_Q^{-1} \left[
                            1+(1-\tau_Q^{-1})f(\tau_Q) \right]
                            g_{QQh/H} \nonumber \\
    F_Q^A (\tau_Q) & = & \tau_Q^{-1} f(\tau_Q) g_{QQA}
\end{eqnarray}
where the coefficients $g_{QQ\Phi}$ denote the couplings of the Higgs bosons
normalized to the $\SM$ Higgs couplings to top and bottom quarks and are given
in Tab.~1. The leading
electroweak radiative corrections are taken into account for the
Higgs masses and couplings. For small values $\tb \sim 1$, the $t$ couplings
are dominant while $b$ couplings are large for large $\tb$. The form factors
are normalized such that for
\begin{displaymath}
\begin{array}{llll}
\mbox{$m_Q \gg M_\Phi$:}\hspace{1cm}
& F^\Phi_Q &\ra& g_{QQ\Phi} \hspace*{6cm} \\
\mbox{$m_Q \ll M_\Phi$:}\hspace{1cm}
& F^{h/H}_Q &\ra& \displaystyle
6 \frac{m^2_Q}{M^2_{h/H}} \left[1 - \frac{1}{4} \left( \log
\frac{M^2_{h/H}}{m^2_Q} - i \pi\right)^2\right] g_{QQh/H}  \\
& F^A_Q &\ra& \displaystyle - \frac{m^2_Q}{M^2_A} \left( \log
\frac{M^2_A}{m^2_Q} - i \pi\right)^2 g_{QQA}
\end{array}
\end{displaymath}
Both form factors approach zero in the chiral limit $m_Q \ra 0$. The leading
logarithmic terms also give rise to the same cross section in the scalar and
pseudoscalar case in the approach to the chiral limit.

The diagrams for the QCD radiative corrections are the same shown in
Figs.~11. As in the $\SM$ case, the renormalization program has been
carried out in the $\overline{\rm MS}$ scheme and the ``physical" quark
mass is defined at the pole of the propagator. The renormalization is also
the same; however for the pseudoscalar case one has to deal with the problem
of $\gamma_5$ in the dimensional regularization scheme and we have adopted the
't Hooft--Veltman scheme.\cite{TV} In this scheme, the renormalization
$Z_{AQQ}$ of the pseudoscalar $AQ \overline{Q}$ vertex must be supplemented
by an additional term $8 \alpha_s / (3 \pi)$ to restore chiral invariance in
the limit $m_Q \ra 0$ and the correct form of the Adler--Bell--Jackiw
anomaly.\cite{ABJ}
As usual, after adding up all contributions, ultraviolet and infrared
singularities cancel and the leftover collinear singularities are absorbed into
the re\-nor\-ma\-lized (in the $\overline{\rm MS}$) parton densities.

The calculation\cite{X7} has been performed following the same lines
as in the Standard Model case and the results for the cross sections are given
by eqs.~(3.12--3.15) with $H$ replaced by $\Phi=h/H/A$. In the limit of large
quark masses, the corrections can be calculated analytically, and one
obtains\cite{X7,N5}
\begin{eqnarray}
m_Q \gg M_\Phi : \hspace{2cm} c^{h/H}_Q \ra 11/2 \ , \hspace*{2cm} c^A_Q \ra 6
\end{eqnarray}
and the coefficients $d^\Phi_{q \overline{q}}$, $d^\Phi_{gq}$ and $d^\Phi_{gg}$
coincide for scalar and pseudoscalar Higgs particles and are given in
eq.~(3.16).\cite{X7,N5} Also in the chiral limit
of small quark but large Higgs masses, these coefficients approach a common
limit for scalar and pseudoscalar Higgs particles.

The $K$ factors which characterize the size of the QCD radiative corrections
properly, are defined by the ratios $K^\Phi_{tot} = \sigma^\Phi_{HO} /
\sigma^\Phi_{LO}$. The cross sections\cite{X7} $\sigma^\Phi_{HO}$ in
next--to--leading order are normalized to $\sigma^\Phi_{LO}$, evaluated for
parton densities and $\alpha_s$ in leading order. $K^\Phi_{tot}$ split into
contributions from the (regularized) virtual corrections $K^\Phi_{virt}$ plus
the real corrections $K^\Phi_{ij} = \Delta \sigma^\Phi_{ij} /
\sigma^\Phi_{LO}$. For both the renormalization and the factorization scale
$\mu = M = M_\Phi$ has been chosen.

The $K$ factors can be determined in the way defined above,
by adopting the GRV parametrizations of the parton
densities\cite{GRV} for which separate $LO$ and $HO$ analyses have been
performed. As shown in Fig.~14, $K^\Phi_{virt}$ and $K^\Phi_{gg}$ are of
similar size and of order 50\%, in general, while
$K^\Phi_{gq}$ and
$K^\Phi_{q \overline{q}}$ are quite small. Apart from the
$t \overline{t}$ threshold region, the $K$ factors
$K^\Phi_{tot}$ are rather
insensitive to the values of the Higgs masses. Near the threshold the
present perturbative analysis, based on one--gluon exchange, cannot
be applied anymore for the pseudoscalar particle $A$. In particular,
since $t \overline{t}$ pairs at rest can form $0^{-+}$ bound states, the
$Agg$ coupling develops a Coulombic singularity for
$M_A \approx 2 m_t$. The range within a few GeV of the threshold
mass must therefore be excluded from the analysis.

Outside this singular range, the QCD corrections are in general large
and positive with values up to $\sim 2$, except for the light
neutral Higgs boson $h$ and small values of $M_A$ for which the $K$
factor is close to unity for
large $\tb$. As expected, the $K$ factor for $h$ coincides with the
corresponding $\SM$ value if $M_h$ appoaches its maximum for a given
$\tb$. [In this limit the $\MSSM$ reduces effectively to the
$\SM$ with one light Higgs boson and $\SM$ type couplings.] The rapid
change of the $K$ factors near this limit corresponds to a rather
gradual change if the $\SUSY$ space is parametrized by $M_A$ instead
of $M_h$.
The $K$ factors do not vary dramatically with the Higgs masses. The
analytical results in the limit $\tau_Q \ra 0$ provide in
general a useful guideline for processes mediated by top loops. \\

\vspace*{11.4cm}

\nn {\small Fig.~14 Individual and total ``K--factors" for the QCD corrections
to the processes $gg \ra h/H/A$ for two values of ${\rm tg}\beta= $ 2.5 and 20;
$m_t=175$ GeV and $M_S=1$ TeV.}  \s

\nn 3.2.3 {\it Higgs Detection} \s

\nn As for the Standard Model Higgs boson, the various signals which can be
used
for the detection of supersymmetric neutral Higgs bosons are: two isolated
photons from $gg \ra h/H/A \ra \gamma \gamma$; a tagged lepton and two
isolated photons $pp \ra l \nu \gamma \gamma$; the ``gold--plated" signal
$pp\ra H\ra ZZ\ra 4l^\pm$; in addition one can use $H,A \ra \tau^+ \tau^-$
mode which is hopeless in the case of the $\SM$ Higgs boson. The charged Higgs
boson can be detected in top quark decays. The situation at hadron colliders
can be summarized as follows:

i) Since the lightest Higgs boson mass is always smaller than $\sim 140 $
GeV, the $ZZ$ signal where one of the $Z$ bosons is off--shell cannot be used
for it.
Furthermore, the $hWW(h\bar{b}b)$ coupling is suppressed (enhanced) leading to
a smaller $\gamma \gamma$ branching ratio than in the $\SM$ [additional
contributions from chargino and sfermion loops can also alter the two--photon
decay width] making the search for the $h$ boson more difficult than for the
Standard Model Higgs boson.

ii) Since the pseudoscalar $A$ has no tree level couplings to gauge
bosons and since the couplings of the heavy $\CP$--even $H$ boson are strongly
suppressed, the gold--plated $ZZ \ra 4l^\pm$ signal is lost [for $H$ it
survives
only for small $\tb$ and $M_H$ values, provided that $M_H<2m_t$]. One has
therefore to rely on the $A,H\ra \tau^+\tau^-$ channels for large $\tb$ values;
this mode, which is hopeless for the $\SM$ Higgs boson, seems to be feasible in
the supersymmetric case. The channel where the neutral Higgsses are produced
with $t\bar{t}$ pairs and decay to $b$ quarks seems also promising\cite{DJV}.

iii) As discussed previously, charged Higgs particles, if lighter than the
top quark, can be accessible in top decays $t \ra H^+b$. This results in a
surplus of $\tau$ lepton final states [in this mass range, the main decay mode
of the charged Higgs is $H^-\ra \tau \nu_\tau$] over $\mu,e$ final states, an
apparent breaking of $\tau \ vs. \ e,\mu$ universality. At LHC, $H^\pm$ masses
up to $\sim 100$ GeV can be probed for $m_t\simeq 150$ GeV.

Note that, if the mass of the lightest Higgs boson $h$ is close to its maximal
allowed value for a given value of $\tb$, all the other Higgs bosons are very
heavy [and degenerate] and $h$ has exactly the couplings of the $\SM$ Higgs
boson. In this case, the situation is similar to the $\SM$ case with $M_H=$
100--140 GeV, unless squark or chargino loops alter the $gg$ production  and
the $\gamma \gamma$ decay processes. Note also that if the neutralinos and
charginos are light enough, Higgs decays into these particles are possible and
would significantly change the search strategies.\cite{Baer} In particular, if
the neutral Higgs bosons can decay into lightest neutralino pairs with large
branching ratios, the search for $\SUSY$ Higgs bosons could be jeopardized at
hadron colliders.

In conclusion, the search for $\SUSY$ Higgs bosons is more difficult than the
search for the $\SM$ Higgs, unless $M_h$ is close to its maximum for a given
$\tb$. Detailed analyses have shown that there is a substantial area in the
$\SUSY$ parameter space where no Higgs particle can be found at the $pp$
collider LHC; this is illustrated in [the hatched area of] Fig.~15 from
Ref.\cite{N1a}

\vspace*{7cm}

\nn {\small Fig.~15 $\SUSY$ parameter space where Higgs bosons can be produced
at the LHC, the top mass was fixed to 150 GeV; from Ref.\cite{N1a}}


\renewcommand{\theequation}{4.\arabic{equation}}
\setcounter{equation}{0}

\section{Production at $\ee$ Colliders}

\nn $\ee$ linear colliders operating in the energy range between 300 to 500 GeV
have a very high physics potential, especially for the discovery of Higgs
particles and the study of their fundamental properties. In this section, we
will survey Higgs physics at these future colliders, and for illustration we
will assume in most of the cases a centre of mass energy of $\sqrt{s}=500$ GeV.

\subsection{Standard Model}

\nn At $\ee$ linear colliders operating in the 300--500 GeV energy range, the
main production mechanisms for Higgs particles, Fig.~16, are the following
processes:
\begin{eqnarray}
{\rm bremsstrahlung \ process} : & & \ee \lra (Z) \lra Z+H \\
{\rm WW \ fusion \ process} : & & \ee \lra \bar{\nu}\ \nu \ (WW)\lra
\bar{\nu}\ \nu \ +H \\
{\rm ZZ \ fusion \ process} : & & \ee \lra e^+ e^- (ZZ) \lra e^+ e^- +
H \\
{\rm radiation \ off \ tops}:& & \ee \lra (Z, \gamma) \lra t \bar{t}
+ H
\end{eqnarray}

\vspace*{5cm}

\nn {\small Fig.~16 Main production mechanisms for $\SM$ Higgs particles in
high-energy $\ee$ colliders.} \\

The bremsstrahl process\cite{E1} is dominant for moderate values of the ratio
$M_H/\sqrt{s}$, but falls off according to the scaling law $\sim s^{-1}$ at
high energies. The fusion processes,\cite{E2} on the other hand, are most
important for
small values of the ratio $M_H/\sqrt{s}$, i.e. high energies where the cross
sections grow $\sim M_W^{-2}$log$s$. For Higgs masses in the intermediate mass
range, the cross sections for the bremsstrahl process and the $WW$ fusion
process are of comparable size at $\sqrt{s}=500$ GeV, while the $ZZ$ fusion
cross section is smaller by an order of magnitude [Fig.~17]. With $\sigma
\sim $ 100 fb, a total of $\sim $ 1000 Higgs particles can be created at an
integrated luminosity of $\int {\cal L} =10\ {\rm fb}^{-1}$, which corresponds
to a running time of $10^{7}$s per year at ${\cal L}= 10^{33} {\rm cm}^{-2}{\rm
s}^{-1}$. For Higgs masses below 100 GeV, the cross section for Higgs radiation
off top quarks, $\ee \ra t\bar{t}H$,\cite{X8} is of the order of a few fb;
this process can be used only to measure the $ttH$ Yukawa coupling once the
Higgs boson is detected in the previous processes.

Additional production mechanisms are provided by $\gamma \gamma$ and
$e \gamma$ collisions,\cite{E3} the high--energy photons generated by Compton
back--scattering of laser light. Higgs particles can then be created in
$\gamma \gamma$ fusion
\beq
\gamma \gamma \lra H
\eeq
and through bremsstrahlung off the $W$ line
\beq
e \gamma & \lra & \nu W H
\eeq

The most important production processes in $\ee$ collisions will be discussed
in some detail below.

\vspace*{9cm}

\nn \begin{center}
{\small Fig.~17 Production cross sections for $\SM$ Higgs particles
at $\sqrt{s}=500$ GeV.}
\end{center}
\vspace*{2mm}

\nn 4.1.1. {\it Higgs Bremsstrahlung} \s

\nn The cross section for the bremsstrahl process eq.~(4.1) can be presented
in a compact form
\beq
\sigma(\ee \ra ZH) = \frac{G_F^2 M_Z^4}{96 \pi s} (v_e^2+a_e^2)
\ \lambda^{1/2} \frac{ \lambda+ 12M_Z^2/s}{(1-M_Z^2/s)^2}
\eeq
where $a_e=-1$ and $v_e=-1+4s_W^2$ are the $Z$ charges of the electron and
$\lambda=(1-M_H^2/s-M_Z^2/s)^2-4M_H^2M_Z^2/s^2$ is the usual two--particle
phase
space function. With $\sigma \sim 200$ fb, a rate of $\sim
2000$ Higgs particles in the intermediate mass range is produced at an energy
$\sqrt{s}=300$ GeV and an integrated luminosity of $\int {\cal L}=10$ fb$^{-1}
$; see Fig.~18. Asymptotically, the cross section scales $\sim 1/s$.

The angular distribution of the $Z/H$ bosons in the bremsstrahl process is
sensitive to the spin of the Higgs particle\cite{X9} as will be discussed
later.
For high energies, the $Z$ boson is produced in a state of longitudinal
polarization so that -- according to the equivalence theorem\cite{E4} -- the
production amplitude becomes equal to the amplitude $A(\ee \ra \phi^0H)$, with
$\phi^0$ being the neutral Goldstone boson which is absorbed to give mass to
the vector boson. The angular distribution, ${\rm d}\sigma/{\rm d}\cos\theta
\sim \lambda \sin^2 \theta + 8M_Z^2/s$, therefore approaches the spin--zero
angular distribution asymptotically

\vspace*{-0.8cm}

\beq
\frac{{\rm d}\sigma}{\sigma {\rm d}\cos \theta} \rightarrow \frac{3}{4}
\sin^2\theta
\eeq
The radiative corrections to the angular distribution and the
total cross section are well under control.\cite{E5} The bulk
of these corrections is due to photon radiation from the incoming electrons and
positrons. The weak corrections are relatively modest being of the order of a
few percent;\cite{PR2} see Fig.~18. \\

\vspace*{6.8cm}

\nn {\small Fig.~18 Total cross sections for the bremsstrahl process for
three energy values $\sqrt{s}=300,400$ and $500$ GeV with and without
electroweak radiative corrections.} \\

The recoiling $Z$ boson in the two--body reaction $\ee \ra ZH$ is
mono--energetic, $E_Z = (s-M_H^2+M_Z^2)/(2\sqrt{s})$, and the mass of the Higgs
boson can be derived from the energy of the $Z$ boson, $M_H^2 =s -2\sqrt{s} E_Z
+M_Z^2$, if the initial $e^+$ and $e^-$ beam energies  are sharp. However,
beamstrahlung smears out the c.m. energy and the system moves along the beam
axes [initial state photon radiation and the beam energy spread have also to be
taken into account]. The intensity of the beamstrahlung depends on the machine
design. It must be suppressed as strongly as possible for a given luminosity in
order to allow for the high quality experimental analyses which are based on
the powerful kinematical constraints familiar from low--energy $\ee$ physics.
Good examples in this context are the DESY--Darmstadt\cite{E6} and the
TESLA\cite{E7} design studies; see Fig.~19. For these designs  the smearing of
the missing mass is of the same magnitude as the experimental uncertainties in
the reconstruction of the $Z$ boson in the leptonic decay channels. This is
shown in Fig.~20a where the missing mass in the signal $\ee \ra ZX, \ X=H$, is
compared for $M_H=130$ GeV with the background from $\ee \ra ZZ$ final states.
\\

\vspace*{6.3cm}

\begin{center} {\small Fig.~19 Beamstrahlung corrections for $\ee \ra ZH$ total
cross sections.\\} \end{center}

Since the recoiling $Z$ boson remains approximately mono--energetic,
even if beamstrahlung is taken into account, it is easy to separate the
signal from the background.\cite{PR4} Four mass regions must be considered:

(i) For Higgs masses close to the $Z$ mass, double $Z$--production
$\ee \ra ZZ$ is the main background source. The cross section is large but can
be reduced by cutting out the forward production and by selecting $b\bar{b}$
final states by means of flavor tagging through vertex detection.
While the Higgs particle decays almost
exclusively to $b\bar{b}$ final states, the branching ratio of the decay
$Z \ra b \bar{b}$ is small, $\sim 15\%$.

(ii) For masses between 100 and 140 GeV, the background comes from
single $Z$--production in $\ee \ra Z Z^*(\ra q \bar{q})$. The cross section is
suppressed by one order of the electroweak coupling compared to the signal.
Further reduction of the background can be achieved through flavor tagging.

(iii) In the mass range above $\sim 140$ GeV where gauge boson decays
become dominant, the most important background is due to $\ee \ra Z+WW^*(\ra q
\bar{q}')$. The cross section of this reaction is suppressed by two
powers of the electroweak coupling relative to the signal.

(iv) Beyond 160 GeV and 180 GeV, the reactions with three gauge
bosons in the final state, $\ee \ra Z+WW$ and $\ee \ra Z+ZZ$ are the main
background channels. In the background the invariant mass of the $WW$ or $ZZ$
final states is broad as opposed to the resonance structure of the signal.
The background under the signal is therefore small as demonstrated in Fig.~20b.

\newpage

\vspace*{15.5cm}

\nn {\small Fig.~20 (a) Distribution of the missing mass $M^2=(p_{e^+}+
p_{e^-}-p_{\mu
^+}-p_{\mu^-})^2$ for the $\ee \ra \mu^+\mu^-H$ signal assuming $M_H=130$~GeV
and the $\ee \ra \mu^+\mu^-Z$ background, including bremsstrahlung,
beamstrahlung and the beam-energy spread [angular cut $|\cos
\theta_{\vec{\mu\mu}}| <0.6]$. Distributions d$\sigma/$d$M_{VV}$ (in pb/GeV)
for heavy--Higgs signals plus backgrounds in the (b) bremsstrahl and (c)
$WW/ZZ$-fusion channels for a selection of $M_H$ values; acceptance cuts are
applied and bremsstrahlung, beamstrahlung and the beam-energy spread are taken
into account. Everywhere the DESY-Darmstadt narrow-band design with $\sqrt
s=500$~GeV has been assumed; the results for the TESLA design are very
similar.}

\newpage

\nn 4.1.2 {\it Fusion Processes} \s

\nn The cross section for the fusion processes can be cast into
a compact form
\beq
\sigma= \frac{G_F^3 M_V^4}{64 \sqrt{2} \pi^3} \int_{\kappa_H}^1 {\rm d}x
\int_x^1 \frac{ {\rm d}y}{[1+(y-x)/ \kappa_V]^2} \left[ (v^2+a^2)^2 f(x,y)
+ 4 v^2 a^2 g(x,y) \right]
\eeq
\vspace*{-6mm}
\beq
f(x,y) &=& \left(\frac{2x}{y^3} -\frac{1+2x}{y^2} +\frac{2+x}{2y} -\frac{1}{2}
\right)\left[ \frac{z}{1+z} -\log (1+z) \right] +\frac{x}{y^3}\frac{z^2(1-y)}
{1+z} \non \\
g(x,y) &=& \left(-\frac{x}{y^2}  +\frac{2+x}{2y} -\frac{1}{2} \right)
\left[ \frac{z}{1+z} -\log (1+z) \right] \non
\eeq
with $\kappa_H =M_H^2/s, \kappa_V=M_V^2/s ,z=y(x-\kappa_H)/(\kappa_V x)$ and
$v, a$ the electron couplings to the massive gauge bosons [$v=-1+4s_W^2 ,
a=-1$ for the $Z$ boson and $v=a=\sqrt{2}$ for the W boson]. \\

\vspace*{12cm}

\nn {\small Fig.~21 Total cross sections for the fusion mechanisms for
three energy values $\sqrt{s}=300,400$ and $500$ GeV as a function of the
Higgs boson mass.} \\

For a total energy $\sqrt{s}=500$ GeV
and a Higgs mass in the intermediate range, the $WW$ fusion cross section is of
about the same magnitude as the bremsstrahl cross section, for lower energies
it is smaller and for higher energies larger [see Fig.~21]. Asymptotically it
keeps growing logarithmically $ M_W^{-2} \log s/ M_H^2$ in contrast to the
bremsstrahl cross section which falls $\sim s^{-1}$. The cross section for $ZZ$
fusion is about an order of magnitude smaller than the cross section for $WW$
fusion; this is a mere consequence of the fact that the NC couplings are
smaller than the CC couplings. The lower rate however is, at least partly,
compensated by the clean signature of the $\ee$ final state in (4.3) that
allows
for a missing--mass analysis to tag the Higgs particle. [The smearing of the
Higgs peak due to beamstrahlung is similar to the case of Higgs bremsstrahlung
in Fig.~20b].

The production of the Higgs particle in the fusion processes is central with
a spread in rapidity of about one unit; the energy distribution peaks at about
30 GeV above the mass value. The transverse momentum of the Higgs particle has
a broad maximum at $P_T \sim 60$ GeV.
For a light Higgs mass, the by far dominant
background is the process $\ee \ra e^+W^-\nu_e$; the cross section  exceeds the
signal for jet--jet final states by about a factor of 60. Another
background arises from $WW$ fusion into a $Z$ boson which is three times larger
than the signal. All other backgrounds can be efficiently reduced.\cite{PR4}

The single $W(Z)$ boson production shows a  behavior similar to the Higgs boson
in the signal process such that cuts enhance the signal/background ratio very
little except for three distinctive features: the resonance structure, the spin
of the resonance and the flavor composition of the decays.\cite{PR4} As the
Higgs mass enters the $Z$ and $W$ resonance region, flavor tagging is
indispensable. Its application would lead to an event sample of about 240
events composed of 60 $W \ra jj$, 80 $Z \ra jj$ and 100 $H \ra jj$ tagged as $b
\bar{b}$--jets [for realistic tagging efficiencies and a luminosity of 20
fb$^{-1}$]. For a Higgs mass around $2M_W$, the background process with
$W^+W^-$ and $WZ$ final states can be reduced to a negligible level.

The process $\ee \ra \ee H$ is, at $M_H \sim M_Z$, contaminated by $\ee \ra
\ee Z$ which is about 10 times stronger than the signal after requiring both
$e^+$ and $e^-$ to be detected with $P_T >30$ GeV but before tagging the $b$'s
in the Higgs decay. The signal to background ratio improves rapidly as $M_H$
moves away from the $Z$ mass. \\

\nn 4.1.3. {\it Measurement of Higgs Couplings} \s

\nn The fundamental particles acquire masses through the interaction with the
Higgs field. The size of the Higgs couplings to fermions and gauge bosons is
therefore set by the masses of these particles. This is a necessary requirement
to unitarize the theory of electroweak interactions. Once the Higgs is found,
it will be of great importance to measure its couplings to the fundamental
particles, which are uniquely predicted by the Higgs mechanism.

The Higgs couplings to massive gauge bosons can directly be
determined from the measurement of the cross sections: the $HZZ$
coupling in the bremsstrahl and in the $ZZ$ fusion processes; the $HWW$
coupling in the $WW$ fusion process.   For sufficiently large Higgs masses
above $\sim$ 250 GeV, these couplings can also be determined experimentally
from the decay widths $H \ra VV$, $V=Z,W$.
Higgs couplings to fermions are not easy to measure directly.
For Higgs bosons in the intermediate mass range where the decays into
$b\bar{b},c \bar{c}$ and $\tau^+ \tau^-$ are important, the decay width is so
narrow that it cannot be resolved experimentally. Nevertheless, the branching
ratios into $\tau$ leptons and charm quarks reveal the couplings of these
fermions relative to the coupling of the $b$ quarks into which the Higgs boson
decays predominantly. In the upper part of the intermediate mass range but
below the threshold for real $WW$ decays, the branching ratios BR($H \ra VV^*$)
are sizeable and can be determined experimentally\cite{Hildreth}. In this case,
the absolute values of the $b$ and eventually of the $c , \tau$ couplings can
be
derived once the $HVV$ couplings are fixed by the cross sections.

The decays $H \ra gg$ and $\gamma\gamma , Z\gamma $ and the process $\gamma
\gamma \ra H$ are mediated by loop diagrams and are proportional to the
couplings of the Higgs boson to heavy particles, not only the top quark but
also particles beyond the $\SM$. Indeed, the number of heavy particles can be
counted in these processes if their masses are generated by the Higgs
mechanism, so that their coupling to the Higgs grow with the mass; one
therefore needs a more direct way to measure the $ttH$ coupling.

A direct way to determine the Yukawa coupling of the intermediate
mass Higgs boson to the	top quark in the range $m_H \le	120$ GeV is
provided by the	bremsstrahl process\cite{X8,X11} $e^+e^- \ra t\overline{t} H$.
For large Higgs masses above the $t\bar{t}$ threshold,	the decay channel $H
\ra t\bar{t}$ increases the cross section of $e^+e^- \ra t\bar{t}Z$ through the
reaction\cite{E8}  $\ee \ra ZH(\ra t\bar{t})$; without the Higgs decay this
final state is produced mainly through virtual $\gamma$ and $Z$ bosons.

The $t\bar{t}H$ final state is generated almost exclusively through Higgs
bremsstrahlung off the top quarks. Additional contributions from Higgs
particles emitted by the $Z$ line, are very small. In general, the top and
Higgs masses must be kept non--zero so that the cross section for Higgs
bremsstrahlung is quite involved. However, neglecting these mass effects
and the Higgs emission from the $Z$ line at a level of ${\cal{O}}(10\%)$, the
Dalitz plot density can be written in a simple form\cite{X8,X11}
\begin{eqnarray}
\frac{{\rm d}\sigma}{{\rm d}x_t {\rm d}x_{\overline{t}}} =
\frac{\alpha^2}{s}  \frac{g^2_{ttH}} {4\pi} \hspace*{-4mm} & & \left\{ \left[
Q_e^2 Q_t^2 + \frac{2 Q_e Q_t \hat{v}_e\hat{v}_t}{1-M_Z^2/s}+\frac{(\hat{v}_e^2
 + \hat{a}_e^2) (\hat{v}_t^2 + \hat{a}_t^2)}{(1- M_Z^2/s)^2} \right] \right.
\non \\
& & \left. \times \frac{x_H^2}{(1-x_t)(1-x_{\overline{t}})}-\frac{(\hat{v}_e^2
+ \hat{a}_e^2) \hat{a}_t^2}{(1 -M_Z^2/s)^2} 2(1+x_H) \right\}
\end{eqnarray}
where $\hat{v}_e,\hat{a}_e$ are the $Z$ charges of the electron
introduced earlier, $x_H = 2E_H/\sqrt{s}$ is the reduced energy of the Higgs
boson and $g_{ttH}$ is the coupling of the Higgs particle to
the top quark.
The integrated cross section is shown for various c.m. energy values
in Fig.~22 as a function of $M_H$. At $\sqrt{s}=500$ GeV, while for small
$M_H$ the cross sections increase with $m_t$ as a result of the rising Yukawa
coupling, this trend is reversed for heavy	Higgsses
by the reduction of the available phase space. For an
integrated luminosity of $\int{\cal{L}}	= 20 \mbox{ fb}^{-1}$, some
100 events can be expected at Higgs masses of order 60 GeV, falling to
less than 20 events at 100 GeV. \\

\vspace*{6.8cm}

\nn {\small Fig.~22 The cross sections $\sigma(e^+e^- \rightarrow t \bar{t}H)$
at $\sqrt{s}=0.5, 1$ and $1.5$ TeV as a function of the Higgs mass; the
top quark mass is fixed to 175 GeV.} \s

Taking acceptance	losses into account, this appears	to be the
upper	limit at which the $ttH$ coupling in the intermediate mass range can be
measured directly in the	course of a few years. Since the signature of
the process $e^+e^- \ra t\overline{t}H \ra WWbbbb$	is spectacular,	there
is reasonable hope to isolate these events experimentally despite the low
rates. The large number of $b$ quarks together	with the mass constraints
$m(bb) = m(H)$ and $m(Wb) = m(t)$ will be crucial in rejecting background
events. QCD initiated $t\overline{t}b\overline{b}$ final states are suppressed
strongly, $\sigma_{QCD} < 0.1$ fb, if the	$b\overline{b}$ invariant mass
is restricted to values larger than 50 GeV.

The reaction $e^+e^- \ra t\overline{t}Z$ is generally
mediated by the subprocess $e^+e^- \ra (\gamma^*,Z^*) + Z$ with  the top
quark	pair coupled to the	virtual	photon and $Z$	boson.  The
Higgs	contribution is important only if the Higgs boson is
produced as a real particle, with a mass large enough to allow for the	decay
into a top quark pair; the effect on the cross section is
most pronounced for Higgs masses not far above the top decay
threshold.\cite{E8} At a 500 GeV collider and because the top mass is large,
there is not enough available phase space and the cross section $\sigma(e^+e^-
\ra Zt \bar{t})$ is extremely small being of the order of 1 fb; the effect of
$H \ra t\bar{t}$ is also marginal; therefore, higher energy colliders will be
required for this process to be useful. Note that for large Higgs masses the
partial width into top quarks rises linearly with $M_H$ while the decay widths
to $W$ and	$Z$ bosons grow	with the third power, thus suppressing the
branching ratio	to top quark pairs.

Even though	both reactions,	Higgs bremsstrahlung off top quarks and
Higgs decay to top quark pairs,	are not	easy to	handle experimentally
in view	of the small cross sections, they nevertheless deserve attention
as they	may provide the	only opportunity to measure the
Higgs--fermion coupling	directly.

\nn 4.1.4 {\it Measurement of spin and parity} \s

\nn The scalar character of the Higgs particle can be tested at $\ee$ colliders
in several ways. The angular distribution of the $\ee \ra ZH$ final state
depends on the spin and parity of the Higgs particle.\cite{X9,E9,ADBK} The
same is true of the angular correlations in the Higgs decay to fermion and
gauge boson pairs.\cite{X9,ADBK,E10}

Since the production of the final state $\ee \ra Z^* \ra ZH$ is mediated by a
virtual $Z$ boson [transversally polarized to the $e^\pm$ beam axis], the
production amplitude could be a monomial in $\cos/\sin \theta$. In the $\SM$,
however, the $ZZH$ coupling, ${\cal L}(ZZH) =
\left(\sqrt2G_F\right)^{1/2}M_Z^2HZ^\mu Z_\mu \ \sim G_F^{1/2}HZ ^{\mu\nu}
Z_{\mu\nu}$, is an S--wave coupling $\sim \vec\epsilon_1\cdot\vec\epsilon_2$ in
the laboratory frame, linear in $\sin \theta$ and even under parity and charge
conjugation, corresponding to the $0^{++}$ assignment of the Higgs quantum
numbers. At high energies the outgoing $Z$ boson is longitudinally polarized
and the angular distribution follows the $\sin^2 \theta$ law, eq.~(4.8).
Nevertheless, it is interesting to confront these properties of the Higgs boson
to those of a pseudoscalar state $A(0^{-+})$ in order to underline the
uniqueness of the $\SM$ prediction. The pseudoscalar case is realized in
two--doublet Higgs models, in which the $ZZA$ couplings are induced by loop
effects.\cite{S2} The effective point-like coupling, ${\cal
L}(ZZA)=(\eta/4)\left(\sqrt2G_F\right)^{1/2}M_Z^2A Z^{\mu\nu}\widetilde
Z_{\mu\nu}$, with $\eta$ being a dimensionless factor and $\widetilde
Z^{\mu\nu}=\epsilon^{\mu\nu\rho\sigma}Z_{\rho\sigma}$, is a P--wave coupling,
odd under parity and even under charge conjugation. It reduces to
$(\vec{\epsilon}_1 \times \vec{\epsilon}_2)\cdot(\vec{p}_1 - \vec{p}_2)$ in the
laboratory frame. Since the $Z$ spins are coupled to a vector, the angular
distribution is again a binomial in $\sin \theta$, d$\sigma/$dcos$\theta \sim
1- \frac{1}{2}\sin^2\theta$, independent of the energy. The total cross section
is proportional to $\beta^3$ characteristically different from $ZH$ production
near threshold.

The angular correlations specific to the S--wave $0^{++}$ Higgs production
in $\ee \ra ZH$ can directly be confronted experimentally with the process $\ee
\ra ZZ$. This process has an angular momentum structure that is distinctly
different from the Higgs process. Mediated by electron exchange in the
$t$--channel, the amplitude is built up by many partial waves, peaking in the
forward/backward direction. The two distributions are compared with each other
in Fig.~23, which demonstrates the specific character of the Higgs production
process.

Since the longitudinal wave function of a vector boson grows with the
energy of the particle [in contrast to the energy independent transverse wave
function] the $Z$ boson in the S--wave Higgs production process (4.1) must
asymptotically be polarized longitudinally. By contrast, the $Z$ bosons
from $ZA$ associate production or $ZZ$ pair production are transversally
polarized [at high energies in the second case]. This pattern can be checked
experimentally. While the distribution of the light fermions in the $Z \ra
f\bar{f}$ rest frame with respect to the flight direction of the $Z$ is given
by $\sin^2\theta_*$ for longitudinally polarized $Z$ bosons, it behaves as
$(1\pm \cos\theta_*)^2$ for transversally polarized states, after averaging
over the azimuthal angles. Including the azimuthal angles, and integrating out
the polar angles, the angular correlations for $\ee \ra ZH$ $[Z\ra f \bar{f}]$
are the familiar $\cos \phi_*$ and $\cos 2 \phi_*$ dependence associated with
${\cal P}$--odd and even amplitudes, respectively,

\begin{eqnarray}
\frac{ {\rm d}\sigma (ZH)} { {\rm d} \phi_*} \sim 1
 -\frac{9\pi^2}{32}\frac{\gamma}{\gamma^2+2}\frac{2v_ea_e}{v_e^2+a_e^2}
\frac{2v_fa_f}{v_f^2+a_f^2} \cos \phi_*+  \frac{1}{2}\frac{1}{\gamma^2+2}
\cos 2\phi_*
\end{eqnarray}
The azimuthal angular dependence disappears for high energies $\sim 1/\gamma$
as a result of the dominating longitudinal polarization of the $Z$ boson. Note
again the characteristic difference to the pseudoscalar $0^{-+}$ case $\ee \ra
ZA$ $[Z \ra f\bar{f}]$, where the azimuthal dependence is ${\cal P}$--even and
independent of the energy in contrast to the $0^{++}$ case; after integrating
out the polar angles,
\begin{eqnarray}
\frac{ {\rm d}\sigma (ZA)} { {\rm d} \phi_*} \sim 1 -\frac{1}{4}
\cos 2\phi_*\,.
\end{eqnarray}
We can thus conclude that the angular analysis of the Higgs production in
$\ee \ra Z^* \ra ZH$ $[Z \ra f \bar{f}]$ allows stringent tests of the ${\cal
J^{PC}}=0^{++}$ quantum numbers of the Higgs boson. This is a direct
consequence of the $0^{++}$ coupling $\epsilon_1\cdot\epsilon_2$ of the $ZZH$
vertex in the production amplitude.

The spin and parity quantum numbers of Higgs particles can also be checked
in the decay processes
\begin{eqnarray}
H \ra V \ V^* \ra (f_1 \bar{f}_2) \ (f_3 \bar{f}_4) \qquad [V=W,Z]
\end{eqnarray}
\nn The angular distributions of the fermions in the decay process are quite
similar to the rules found for the production process; a complete discussion
can be found in Refs.\cite{X9} Note that the angular correlations in
Higgs decays can also be studied at hadron colliders, although background
problems require a more sophisticated discussion.

\vspace*{7cm}

\nn {\small Fig.~23 Angular distributions of the processes $e^+ e^- \ra ZH,
ZA$ and $ZZ$ at $\sqrt{s}=500$ GeV. The Higgs mass is fixed to $M_H=120$ GeV.}

\subsection{Supersymmetric Extension}

\nn 4.2.1 {\it Neutral Higgs Bosons} \s

\nn The main production mechanisms of neutral Higgs bosons at $\ee$
colliders are the bremsstrahl process and pair production,
\begin{eqnarray}
(a) \ \ {\rm bremsstrahlung} \hspace{1cm} \ee & \ra &  (Z) \ra Z+h/H
\hspace{4cm} \\
(b) \ \ {\rm pair \ production} \hspace{1cm} \ee & \ra & (Z) \ra A+h/H
\end{eqnarray}
as well as the fusion processes, similarly to the $\SM$ Higgs boson,
\begin{eqnarray}
(c) \ \ {\rm fusion \ processes} \hspace{0.8cm} \ \ee & \ra &  \nu \ \bar{\nu}
\ (WW) \ra \nu \ \bar{\nu} \ + h/H \hspace{2.3cm} \nonumber  \\
\ee & \ra &  \ee (ZZ) \ra \ee + h/H
\end{eqnarray}
The ${\cal CP}$--odd Higgs boson $A$ cannot be produced in fusion processes
to leading order. \\

\nn The cross sections for the four bremsstrahl and pair production
processes can be expressed as\cite{W6a}
\begin{eqnarray}
\sigma(\ee \ra Zh) & =& \sin^2(\beta-\alpha) \ \sigma_{\rm SM} \nonumber \\
\sigma(\ee \ra ZH) & =& \cos^2(\beta-\alpha) \ \sigma_{\rm SM} \nonumber \\
\sigma(\ee \ra Ah) & =& \cos^2(\beta-\alpha) \ \sigma_{\rm SM} \ \bar{\lambda}
\nonumber \\
\sigma(\ee \ra AH) & =& \sin^2(\beta-\alpha) \ \sigma_{\rm SM} \ \bar{\lambda}
\end{eqnarray}

\nn where
\begin{eqnarray}
\sigma_{\rm SM} = \frac{G_F^2 M_Z^4}{96 \pi s} (v_e^2+a_e^2) \lambda^{1/2}
(M_{h,H}^2,M_Z^2;s) \ \frac{ \lambda(M_{h,H}^2,M_Z^2;s)+12M_Z^2/s}
{(1-M_{Z}^{2}/s)^2}
\end{eqnarray}
\nn is the $\SM$ cross section for Higgs bremsstrahlung and the factor
$\bar{\lambda}$ accounts for the correct suppression of the $P$--wave
cross sections near the threshold
\begin{eqnarray}
\overline{\lambda} \ = \ \frac{\lambda^{3/2}(M_{j}^2,M_A^2; s)}{\lambda^{1/2}
(M_{j}^2,M_Z^2; s) \ [\lambda(M_{j}^2,M_A^2; s)+12M_Z^2/s] } \hspace*{1cm}
[ \ j=h,H \ ]
\end{eqnarray}

\nn The cross sections for the bremsstrahlung and the pair production as
well as the cross sections for the production of the light
and the heavy neutral Higgs bosons $h$ and $H$ are mutually complementary
to each other, coming either with a coefficient $\sin^2(\beta-\alpha)$
or $\cos^2(\beta-\alpha)$. Since $\sigma_{\rm SM}$ is large, at least one of
the
${\cal CP}$--even Higgs bosons should be detected. From the mass and
$\sin^2$/$\cos^2(\beta-\alpha)$ plots, we conclude that the following final
states  will be observed [Fig.~24], depending on the values of the parameters
$M_h$ and $\tb$:
\begin{eqnarray}
\begin{array}{ccccccc}
M_h  \ ``{\rm small}" \ \ , \ \ & \tb \ {\rm small} \ : & \ \ hZ \ ,  &  \ \
& HZ \ , \ \ & hA \ ,  \ \ & HA  \nonumber \\
\vspace*{2mm}
& \tb \ {\rm large} \ : & & &  HZ \ , \ \  & hA \ , \ \ & \nonumber \\
M_h  \ ``{\rm large}" \ \ , \ \ & \tb \ {\rm small} \ : & \ \ hZ \ , & & & &
[HA] \nonumber \\
& \tb \ {\rm large} \ : & \ \ hZ \ , & & & & [HA] \nonumber
\end{array}
\end{eqnarray}

\nn where ``$M_h$ small" and ``large" are synonymous for ``considerably below"
and ``close to the upper limit of the light ${\cal CP}$--even Higgs boson"
for a given value of $\tb$. If $M_h$ is ``large" the $H,A$ masses can exceed
the kinematical limit for $HA$ pair production. \\

\vspace*{10.5cm}

\nn {\small Fig.~24 Regions of the $[M_h,\tb]$ plane where the four cross
sections $\ee \ra hZ, hA, HZ$ and $HA$ are observable. The contours are defined
such that the cross sections are larger than 2.5 fb [corresponding to 25 events
for an integrated  luminosity of $\int {\cal L} =10$ fb$^{-1}$]. The dashed
area corresponds to the theoretically forbidden region [$M_h^2 \not> M_Z^2
\cos^2 2\beta+ \epsilon \sin^2 \beta$]. The region which can be probed at
LEP200 [defined such that for $\sqrt{s}=180$ GeV and $\int {\cal L} =500$
pb$^{-1}$, the number of events in one of the two processes  $\ee \ra hZ$ or
$hA$ is larger than 25] is the area to the left of the thin line. We have taken
$m_t =175$ GeV and $M_S=1$ TeV. The process $\ee \ra hZ$ is accessible in the
\underline{entire area below} the full line, $hA$ in the \underline{entire area
above} the broken line and $HZ$ in the \underline{entire area above} the full
line; $HA$ final states can be detected in the \underline{area between} the two
dashed lines.} \\

\nn The cross sections for the production of the ${\cal CP}$--even light and
heavy Higgs bosons $h$ and $H$ via bremsstrahlung are shown as functions of the
Higgs mass in Fig.~25a. The cross section for $h$ is large for small values of
$\tb$ and/or large values of $M_h$ where $\sin^2(\beta- \alpha)$ approaches its
maximal value.	In these two cases the cross section is	of the order of
$\sim 50$ fb, which  for an integrated luminosity of 10 fb$^{-1}$ corresponds
to $\sim$ 500 events.	By contrast, the cross section for $H$ is
large for large $\tb$ and light $h$ [implying small $M_H$]. In the case of
$h$ [and also for $H$ in most of the parameter space] the signal consists
of a $Z$ boson accompanied by a $b\bar{b}$ or a $\tau^+ \tau^-$ pair.
The signal is easy to separate from the background which comes mainly
from $ZZ$ production if the Higgs mass is close to $M_Z$.

\vspace*{15cm}

\nn {\small Fig.~25 Production cross sections of the ${\cal CP}$--even neutral
Higgs bosons at $\sqrt{s}=500$ GeV as functions of their masses for three
values of ${\rm tg}\beta=2.5,5$ and 20; (a) Bremsstrahl processes $e^+e^-
\rightarrow Z+h/H$, and (b) in association with the pseudoscalar Higgs boson
$e^+e^- \rightarrow A+h/H$.}

\newpage

\nn The cross sections for the associate channels $\ee \rightarrow Ah$ and $AH$
are displayed in Fig.~25b. As anticipated, the situation is opposite to the
previous case: the cross section for $Ah$ is large for light $h$ and/or
large values of $\tb$ whereas $AH$ production is preferred in the complementary
region. The sum of the two cross sections decreases from $\sim 50$ to 10 fb
if $M_A$ increases from $\sim 50$ to 200 GeV. In major parts of the parameter
space, the signals consist of four $b$ quarks in the final state, requiring
facilities for efficient $b$ quark tagging. Mass constraints will help to
eliminate the backgrounds from QCD jets as well as $ZZ$ final states. \\

\vspace*{14cm}

\nn {\small Fig.~26 Production cross sections of the ${\cal CP}$--even neutral
Higgs bosons at $\sqrt{s}=500$ GeV as functions of their masses for three
values of ${\rm tg}\beta=2.5,5$ and 20; (a) $WW$ fusion process $e^+e^- \ra
\nu \bar{\nu} +h/H$ (b) and  $ZZ$ fusion process $e^+e^- \ra e^+e^-+h/H$.}

\newpage

\nn $WW$ and $ZZ$ fusion provide additional mechanisms for the production
of the ${\cal CP}$--even neutral Higgs bosons. They lead to Higgs bosons in
association with a $\nu \bar{\nu}$ or $\ee$ pair in the final state. The cross
sections can again be expressed in terms of the corresponding $\SM$ cross
sections, given in eq.~(4.9)
\begin{eqnarray}
\sigma( \ee \ra (VV) \ra h) & = & \sin^2 (\beta-\alpha) \sigma_{\rm SM}^{VV}
\nonumber \\
\sigma( \ee \ra (VV) \ra H) & = & \cos^2 (\beta-\alpha) \sigma_{\rm SM}^{VV}
\end{eqnarray}
For the $WW$ fusion mechanism, the cross sections are larger than
for the bremsstrahl process if the Higgs mass is moderately small -- less than
160 GeV at $\sqrt{s} = 500$ GeV. However, since the final state cannot be fully
reconstructed, the signal is more difficult to extract. As in the case of the
bremsstrahl process, the production of light $h$ and heavy $H$ bosons are
complementary. The cross sections for the $ZZ$ fusion mechanism are about an
order of magnitude smaller than for the one for $WW$. $ZZ$ fusion will
nevertheless be useful as the final state can be fully reconstructed.
The cross sections are displayed for representative values of $\tb$ in
Fig.~26.

The preceding discussion on the neutral $\MSSM$ Higgs sector in $\ee$
linear colliders can be summarized in the following points:\cite{X5,W1}

(i) The lightest ${\cal CP}$--even Higgs particle $h$ can be
detected in the entire range of the $\MSSM$ parameter space, either through the
bremsstrahlung process	$\ee \rightarrow hZ$ or through pair production $\ee
\rightarrow hA$. In fact, this conclusion holds true even at a c.m. energy of
300 GeV, independently of the top and squark mass values, and also if invisible
neutralino decays are allowed for.

(ii) There is a subtantial area of	the [$ M_h,\tb$] parameter
space where {\it all} neutral $\SUSY$ Higgs bosons can be discovered at a
500 GeV
collider.  This is possible if the masses of the scalar $H$ and the
pseudoscalar $A$ boson are less than $\sim 230$ GeV.

(iii) In some part of the $\MSSM$ parameter space, the lightest Higgs
particle $h$ can be detected, but it cannot be distinguished from the $\SM$
Higgs boson [if $\SUSY$ decays are not allowed]. This happens if, for a given
value of $\tb$, $M_h$ is very close to its maximum value and $H$ and $A$ are
too
heavy to be produced in association. In this case, the couplings of $h$ to
gauge bosons and fermions are $\SM$ like. A way to distinguish between the two
is provided by $\ga$ collisons as will be discussed shortly.

(iv) Higgs boson decays into charginos and neutralinos can be very
important in some areas of the $\MSSM$ parameter space; in particular,
invisible
decays of the neutral Higgs bosons can be larger than the decays into
standard particles. At $\ee$ colliders, missing mass techniques allow to
isolate these events in
the bremsstrahlung process for the ${\cal CP}$--even Higgs bosons or in a
mixture of visible and invisible decay modes of $Ah$ and $AH$ in the pair
production processes.

\vspace*{2mm}

Some of these features are not specific to the minimal extension but they are
expected to be realized also in more general $\SUSY$ models. For example, a
light Higgs boson with a mass in the intermediate range is quite generally
predicted in supersymmetric theories.\cite{S3,S4} [The detection of this
particle however is not necessarily guaranteed as the production rate may be
suppressed.]

\nn 4.2.2 {\it Charged Higgs Bosons} \s

\nn An unambiguous signal of an extended Higgs sector would	be the
discovery of a charged Higgs boson.  In	a general two--Higgs doublet
model, charged	Higgs bosons can be as	light as $\sim $ 45 GeV, the
lower limit derived from the negative search at LEP100.\cite{R8}  In the
$\MSSM$ however, $H^\pm$ is constrained to be heavier than the $W$ boson.
More precisely, the lower limit $M_A >45$ GeV obtained at LEP100\cite{AX1}
implies $M_{H^\pm} > 90$ GeV.

In $\ee$ collisions, the production of a pair of charged Higgs
bosons\cite{W6c} proceeds through virtual photon and $Z$ boson exchange.
The cross
section depends only on the charged Higgs mass [and does not depend on
any extra parameter],
\begin{eqnarray}
\sigma(	\ee \longrightarrow H^{+}H^{-}) = \frac{\pi \alpha^2}{3s}
\left[ 1- \frac{ 2 \hat{v}_e \hat{v}_H}{1-M_Z^2/s} + \frac{(\hat{a}_e^2
+\hat{v}_e^2)\hat{v}_H^2} {(1-M_Z^2/s)^2}  \right] \ \beta^3
\end{eqnarray}
with the standard $Z$ charges $\hat{v}_e=(-1+4s_W^2)/4c_Ws_W$, $\hat{a}_e
=-1/4c_Ws_W$ and $\hat{v}_H=(-1+2s_W^2)/2c_Ws_W$, and $\beta=(1-4M_{H^\pm}^2/s
)^{1/2}$ the velocity of the Higgs particles.
The cross section is shown in Fig.~27a as a function of the charged Higgs mass
for a c.m. energy $\sqrt{s}= 500$ GeV. For small Higgs masses the cross section
is of order 100 fb, but it drops very quickly due to the P--wave suppression
factor $\beta^3$ near the threshold. For $M_{H^\pm}=$ 220 GeV, the cross
section has fallen to a level of $ \simeq $ 5 fb, which for an integrated
luminosity of 10 fb$^{-1}$ corresponds to 50 events.  The
angular	distribution of the charged Higgs bosons follows the $\sin^2 \theta$
law typical for spin--zero particle production.

Charged Higgs particles can	also be	created in $\gamma \gamma$ collisions.
Generating the $\gamma$ beams through back--scattering of laser light, the
total energy of the $\gamma \gamma$ collider can go up to $\sim
80\%$ of the original $\ee$ energy, which corresponds to $\sqrt{s_{\gamma
\gamma}} \simeq 400$ GeV	for a 500 GeV $\ee$ collider. The $\gamma
\gamma$ luminosity is expected to be of the same magnitude as the original
$\ee$ luminosity. The cross section is given by
\begin{eqnarray}
\sigma (\gamma \gamma \ra H^+ H^-) = \frac{2\pi \alpha^2}{s}
\beta \left[ 2 -\beta^2 - \frac{1-\beta^4}{2\beta} {\rm log}
\frac{ 1+\beta}{1-\beta} \right]
\end{eqnarray}
where $\beta$ is the velocity of the Higgs particle. The numerical result
is displayed in Fig.~27a  for the $\gamma \gamma$ luminosity without beam
polarization.\cite{WPC,WBBC} Due to the reduced energy, the maximum Higgs mass
which can be probed in $\gamma \gamma$ collisions is smaller than in the
original $\ee$ collisions; the cross section however is enhanced by a factor
$\sim 3$ in the low mass range.

The	charged	Higgs boson, if	lighter	than the top quark, can also be
produced in top decays.\cite{W6b} The ratio of the decay widths $t
\rightarrow bH^+$ and the standard mode $t \rightarrow bW^+$ is given by
\begin{eqnarray}
\frac{\Gamma(t \rightarrow bH^+)}{\Gamma(t \rightarrow bW^+)} &= &
\frac{(m_t^2+m_b^2-M_{H^\pm}^2)(m_t^2{\rm ctg}^2\beta + m_b^2 {\rm tg}^2\beta)
+4m_t^2 m_b^2 }{M_W^2(m_t^2+m_b^2-2M_W^2)+ (m_t^2-m_b^2)^2} \non \\
&& \times \frac{\lambda(M_{H^\pm}^2,m_b^2;m_t^2)^{1/2}}
{\lambda(M_W^2,m_b^2;m_t^2)^{1/2}}
\end{eqnarray}
In the range $ 1 < \tb < m_t/m_b$ favored by $\SUSY$ models, the  branching
ratio BR$(t \rightarrow bH^+)$ varies between $ \sim 2 \%$ and $20 \%$.  Since
the cross section for top pair production is of order of 0.5 pb at $\sqrt{s}=
500$ GeV, this corresponds to 200 and 2000 charged Higgs bosons at a luminosity
$\int {\cal L} =$ 10 fb$^{-1}$; Fig.~27b.

The signature for $H^\pm$ production can be read off the graphs
displaying the branching ratios in section~2. If $M_{H^\pm}
< m_t+m_b$, the charged Higgs boson will decay mainly into $\tau \nu_\tau$
and $c\bar{s} $ pairs, the $\tau \nu_\tau$	mode dominating for $\tb$ larger
than
unity. This results in a surplus of $\tau$ final states over $e, \mu$
final states, an apparent breaking of $\tau \ vs. \ e,\mu$ universality.
For large Higgs masses the dominant decay mode is the top decay $H^+
\rightarrow t \bar{b}$. In some part of the parameter space also the
decay $H^+ \rightarrow  W^+h $ is allowed, leading to cascades with heavy
$\tau$ and $b$ particles in the final state.

\vspace*{13cm}

\nn {\small Fig.~27 Production cross sections of the charged Higgs boson (a)
at $e^+ e^-$ colliders with $\sqrt{s}=500$ GeV and in $\gamma \gamma$
collisions
with $\sqrt{s}=400$ and (b) in top decays with $m_t=175$ GeV and for three
values of $\tb=2.5,5$ and 20.}

\newpage

\subsection{$\gamma \gamma$ Collisions}

\nn Future high--energy $\ee$ linear colliders can be made to run in the $e
\gamma$ or $\gamma \gamma$ mode by using Compton back scattering of laser
light.\cite{WPC,WBBC} One then obtains $e \gamma$ colliders by converting
only one of the initial electron or positron beams to a very energetic photon,
or $\gamma \gamma$ colliders by converting both the electron and positron to
photons. These colliders will have practically the same energy [$\sim 90\%$ for
$e\gamma$ and $\sim 80\%$ for $\gamma \gamma$ machines] and luminosity as the
original $\ee$ collider [although they will depend in a sensitive way on
various machine paramerts]; for details see Ref.\cite{WPC,WBBC}

One of the best motivations for turning single--pass future high--energy $\ee$
linear colliders into $\gamma \gamma$ colliders is undoubtedly the study of the
properties of neutral Higgs particles which can be produced as resonances in
the
$s$--channel. In the context of the supersymmetric and Standard Model Higgs
sector, two main features which are difficult to study in the $\ee$ mode, can
be investigated at such colliders:

$i)$ As discussed previously, since the photons couple to Higgs bosons via
heavy particle loops, the $H\ga$ amplitudes are sensitive to particle masses,
standard and also supersymmetric, well above the Higgs masses themselves. We
have seen in the preceding section that in the $\ee$ mode, the lightest
supersymmetric Higgs particle $h$ can be detected but it cannot be
distinguished from the $\SM$ Higgs boson in some part of the $\MSSM$ parameter
space if $\SUSY$ decays are not allowed. As long as this ambiguity cannot be
resolved by proceeding to higher $\ee$ collider energies, the only way to
distinguish $h$ from the $\SM$ Higgs particle is provided by Higgs production
in $\gamma \gamma$ fusion. While in the Standard Model this process is
built up by $W$ and top quark loops, additional contributions in $\SUSY$
models are provided by supersymmetric particle loops [chargino, sfermion and
charged Higgs boson loops] which alter the $\SM$ production rates.

$ii)$ In the $\ee$ mode, since the standard Higgs boson and the scalar $\MSSM$
Higgs particles $h,H$ couple to vector bosons directly, the positive parity can
be checked by analyzing the $Z$ final states in $\ee \ra Z^* \ra ZH\ [Z \ra f
\bar{f}]$ as discussed above; this method is equivalent to the analysis of the
$h,H \ra VV$ decays. However, in the $\MSSM$ the pseudoscalar $A$ boson has no
tree level couplings to the vector bosons and the latter methods cannot be
used. To study\cite{Wggspin} the ${\cal CP}$ properties of scalar and
pseudoscalar Higgs particles on the same footing [and to check whether ${\cal
CP}$ is violated in the Higgs sector, as it would be the case in a general
two--Higgs doublet model where the three neutral Higgs bosons could correspond
to arbitrary mixed ${\cal CP}$ states], one can run in the $\ga$ mode where
both type of particles are produced through loop diagrams with similar rates.
Indeed, the fusion of Higgs particles by linearly polarized photon beams
depends on the angle between the polarization vectors.\cite{WR24} For scalar
particles the production amplitude $\sim \vec{\epsilon}_1\cdot
\vec{\epsilon}_2$ is maximal only for parallel vectors while pseudoscalar
particles with amplitudes $\sim\vec{\epsilon}_1 \times \vec{\epsilon}_2$
prefer perpendicular polarization vectors. However, for typical experimental
set--ups for Compton back--scattering of laser light,\cite{WPC,WBBC} the
maximum
degree of linear polarization of the generated hard photon beams is less than
about 30\% so that the efficiency for two polarized beams is reduced to less
than 10\%.\cite{WRA2}

To study both features, very high luminosities and a very good control on the
photon beam polarization are required. Moreover, in case where the Higgs bosons
decay into $b$--quarks, one needs a careful analysis of background rejection
due to the enormous number of $b\bar b$ pairs [and also other jets] produced
in $\gamma \gamma$ collisions.\cite{WRA3} For heavy neutral ${\cal CP}$--even
Higgs bosons which decay mostly into massive vector bosons pairs, one also
needs to deal with the large backgrounds from $WW$ and $ZZ$ pair production
[although the latter proceeds only through loop diagrams\cite{WggZZ} as for
Higgs production].

Furthermore, it is mandatory to have a precise prediction for the value of
the Higgs--$\ga$ coupling, especially if one attempts to exploit this coupling
to look for new particles whose masses are not entirely generated through the
Higgs mechanism and therefore have small effects [as is the case for
charginos, sfermions and charged Higgs bosons]. Therefore, one needs to control
properly the radiative corrections to the coupling mediated by the standard
particles and to include the QCD corrections to the quark amplitude. [Note that
a precise prediction of the $\Phi \ra \ga$ decay widths is also important
because they play a crucial r\^ole for the search of these particles in the
lower part of the intermediate mass range at the LHC, as discussed in the
previous section]. For completeness, the QCD corrected $\Phi \ra \ga$ decay
widths will be briefly discussed below.

Denoting the quark amplitudes by $A_Q$ [where $Q=t,b$ in the $\MSSM$] $etc$.,
the ${\cal H}=H_{\rm SM}/ h/H \ra \ga$ and $A \ra \ga$ decay rates are given by
\begin{eqnarray}
\Gamma(\X \ra \ga) &=& \frac{G_F \alpha^2 M_{\X}^3 } {128 \sqrt{2} \pi^3}
\left| \sum_Q N_C e_Q^2 g_{QQ\X} A_Q^\X + g_W^\X A_W^\X
+ \ \cdots \ \right|^2 \\
\Gamma(A \ra \ga) &=& \frac{G_F \alpha^2 M_A^3 }{32 \sqrt{2} \pi^3}
\left| \sum_Q N_C e_Q^2 g_{QQA} A_Q^A + \ \cdots \ \right|^2
\end{eqnarray}
\nn where the quark amplitudes at lowest order are given in eq.~(3.18)
while the $W$ amplitude is given by
\begin{eqnarray}
A_{WW\X} = - \frac{3}{4} \tau^{-1} \left[ 3+2\tau+ 3(2-\tau^{-1})f(\tau) \
\right]
\end{eqnarray}
\nn where, as usual, the scaling variable is defined as $\tau=M_\Phi^2/4m_i^2$
with $m_i$ denoting the loop mass, and the scalar triangle integral $f$ is
given
by eq.~(3.9).

The cross section for the $\ga$ fusion of Higgs bosons is found by folding the
parton cross section with the $\ga$ luminosity, see e.g.~Ref.\cite{WR2A} and
the parton cross section is determined by the $\ga$ width [$s_{\ga}$ is the
photon c.m. energy],
\beq
\sigma( \ga \ra \Phi ) = \frac{8 \pi^2}{M_{\Phi}^3} \Gamma(\Phi \ra \ga) \delta
(1- M_\Phi^2/s_{\ga})
\eeq
The QCD corrections to the quark amplitudes can be parametrized as
\begin{eqnarray}
A_{QQ\Phi} \ = \ A_{QQ\Phi}^{\rm LO} \ \left[ 1+C\frac{\alpha_s}{\pi} \right]
\end{eqnarray}
The coefficient $C$ depends on $\tau=M_\Phi^2/4m_Q^2(\mu^2)$ where the {\it
running} quark mass $m_Q(\mu^2)$ is defined at the renormalization point $\mu$
which is taken to be $\mu=M_\Phi/2$ in our analysis; this value is related
to the pole mass $m_Q(m_Q^2)=m_Q$ in the on--shell renormalization
scheme by
\begin{equation}
m_Q(M_\Phi^2/4) = \displaystyle m_Q  \left[\frac{\alpha_s([M_\Phi/2]^2)}
{\alpha_s(m_Q^2)} \right]^{12/(33-2N_F)} \left\{1+{\cal O}(\alpha_s^2) \right\}
\end{equation}
The lowest order amplitude $A_{QQ\Phi}^{LO}$ is to be evaluated for the
same mass value $m_Q([M_\Phi/2]^2)$. The choice $\mu =M_\Phi/2$ of the
renormalization point avoids large logarithms $\log M_\Phi^2/m_Q^2$ in the
final results for Higgs masses much larger than the quark mass. $\alpha_s$ is
taken at $\mu$ for $\Lambda=200$ MeV.

The two--loop diagrams contributing to the decay widths are a subset of the
diagrams which appear in the case of the $gg \ra$ Higgs fusion processes. We
have evaluated these diagrams in a way similar to what has been discussed in
the $gg \ra H$ case, but for the the running quark mass at $\mu=M_\Phi/2$. The
output of the calculation is shown in Fig.~28 where the amplitudes $C_\X$ for
scalar loops and $C_A$ for pseudoscalar loops are shown as functions of $\tau$.

The coefficients are real below the quark threshold $\tau <1$, and complex
above. Very close to the threshold, within a margin of a few GeV, the present
perturbative analysis cannot be applied anymore. [It may account to some extent
for resonance effects in a global way.] Since $Q\bar{Q}$  pairs cannot form
$0^{++}$ states at the threshold, ${\cal I}mC_\X$ vanishes there; ${\cal
R}eC_\X$ develops a maximum very close to the threshold. By contrast, since
$Q\bar{Q}$ pairs do form $0^{-+}$ states, the imaginary part ${\cal I}mC_A$
develops a step which is built up by the Coulombic gluon exchange [familiar
from the Sommerfeld singularity of the QCD correction to $Q\bar{Q}$ production
in $e^+e^-$ annihilation]; ${\cal R}eC_A$ is singular at the threshold. The
threshold effects have been discussed in detail recently.\cite{Wthresh} For
large $\tau$, both coefficients approach a common numerical value, as expected
from chiral invariance in this limit. In the opposite limit, the QCD
corrections can be evaluated analytically,
\begin{eqnarray}
m_Q \gg M_\Phi: \ \ C_{\X} \ \ra \ - 1 \ \ \mbox{\rm and}
\ \ \ C_A \ \ra \ \ 0
\end{eqnarray}
\nn These results can easily be traced back to the form of the $\ga$ anomaly
in the trace of the energy--momentum tensor\cite{WR11} and to the
non-renormalization of the axial--vector anomaly.\cite{ABJ}
In Fig.~29, the QCD corrected $\ga$ widths for $h,H,A$ Higgs bosons are
displayed in the $\MSSM$ for two values tg$\beta=2.5$ and tg$\beta=20$, taking
into account only quark and $W$ boson loops. While in the first case top loops
give a significant contribution, bottom loops are the dominant component for
large tg$\beta$. The overall QCD corrections are also shown: since we have
used the running quark masses, the corrections to the widths are small, $\sim
{\cal O}(\alpha_s/ \pi)$ everywhere. However,
artificially large $\delta$ values occur for specific large Higgs masses
when the lowest order amplitudes vanish accidentially as a consequence of the
destructive interference between $W$ and quark--loop amplitudes, see also
Ref.\cite{Melnikov}

\newpage

\vspace*{5.5cm}

\nn {\small Fig.~28 Real and imaginary parts of the QCD correction to the
quark amplitudes of the scalar and pseudoscalar couplings normalized to the
Born terms.}

\vspace*{11cm}

\nn {\small Fig.~29 QCD corrected two--photon decay widths [in keV] and size
of the QCD corrections for the $\MSSM$ neutral Higgs bosons with $\tb=2.5$ and
20; the top mass was taken to be $m_t=140$ GeV.}

\newpage

\section{Summary}

\nn Probing electroweak symmetry breaking will be the most important mission
of future
high--energy colliders. In this review, we have discussed the properties of
the Higgs particles of the Standard Model and of its Minimal Supersymmetric
Extension. We have updated various results for couplings, decay widths and
production cross sections to take into account the new value of the top quark
obtained recently by the CDF and LEP/SLC Collaborations, and to include
radiative corrections, some of which have been calculated only recently.

We have then discussed the potential of future high--energy colliders for
discovering and studying the properties of these Higgs particles, and
considered
the case of the CERN hadron collider LHC with a center of mass energy of $\sim
14$ TeV and of a future $\ee$ collider in the energy range 300--500 GeV.

At the hadron collider LHC, Higgs particles can be produced with very high
rates. The main production mechanism is the fusion mechanism $gg \ra H$ which
proceeds through heavy quark loops, followed by the $WW/ZZ$ fusion process and
the associate production with $W/Z$ bosons or heavy quark pairs. The QCD
corrections are important for the $gg$ fusion mechanism, their inclusion
enhances the production rate by practically a factor of two.

For Standard Model Higgs bosons in the ``high mass" region, $M_H>140$~GeV, the
signal consists of the so--called ``gold--plated" events $H \ra Z Z^{(*)} \ra
4l^\pm$ with $l=e,\mu$, with which one can probe Higgs masses up to $\sim 800$
GeV with a luminosity $\int {\cal L}= 100 $~fb$^{-1}$ at LHC. The $H
\ra l^+l^- \bar{\nu} \nu$ and $H \ra WW \ra l \nu jj$ decay channels can push
this limit up to $\sim 1$ TeV. For the ``low mass" range, the situation is
more complicated and one has to rely on the rare $H \ra \gamma \gamma$ decay
mode. At the LHC, the event rate is rather large, ${\cal O}(10^{3})$ events for
$\int {\cal L}= 100$~fb$^{-1}$, but the backgrounds pose a formidable task.
With
a dedicated detector and a high--luminosity option, this channel although very
difficult, is feasible. A complementary channel in this low mass range would be
the $ pp \ra WH, t\bar{t}H \ra \gamma \gamma l \nu$ but unfortunately the rates
are small; a promising process would be $ pp \ra t\bar{t}H$ with the Higgs
decay $H \ra b\bar{b}$, if very high efficiency and purity for $b$--quark
tagging in the LHC environment [especially with the high--luminosity option]
could be made available.

In the Minimal Supersymmetric Standard Model, the situation is even more
complicated. The production mechanisms are the same as for the $\SM$ Higgs but
one has to take the $b$ quark contributions into account. Since the lightest
Higgs boson mass is always smaller than $\sim 140 $ GeV, only the two--photon
decay channel can be used for this particle; the branching ratio is smaller
than in the $\SM$ making the detection of this Higgs boson more difficult,
except in the case where $M_h$ is close to its maximal value where the
situation is similar to the $\SM$ case. Since the pseudoscalar $A$ has no
tree level couplings to gauge bosons and since the couplings of the heavy
${\cal CP}$--even $H$ are strongly suppressed, the gold--plated $ZZ$ signal is
lost in a large part of the parameter space; a promising alternative is to use
the $A,H\ra \tau^+ \tau^-$ channels for large tg$\beta$ values. Another
possibility would be neutral Higgs production in association with $t\bar{t}$
pairs with the Higgs particle decaying into $b$--quarks if very good
micro--vertex detectors are available. Charged Higgs particles, if lighter
than the top quark, can be accessible in top decays. Up to now, there is still
a substantial area in the $\SUSY$ parameter space where no Higgs particle can
be found at the LHC [although, these analyses have to be updated to take into
account the high value of the top quark mass and the new possibilities
offered by $b$--tagging at hadron colliders].

$\ee$ linear colliders with energies in the range 300 -- 500 GeV and a
luminosity of ${\cal L}=$ a few times $10^{33}$cm$^{-2}$s$^{-1}$ are ideal
instruments to search for Higgs bosons in the mass range below the scale of
electroweak symmetry breaking [in fact, standard Higgs bosons can be discovered
up to masses of the order of 350 GeV at a 500 GeV $\ee$ collider]. This range
is the most likely region for the mass of the Standard Model Higgs boson since
in this case, the particle remains weakly interacting up to the GUT scale.

In the intermediate  mass range, Standard Higgs particles can be observed
irrespectively of their decay modes, in three independent production channels:
the bremsstrahlung process $\ee \ra ZH$ and the fusion processes $\ee \ra
\bar{\nu}\nu H$ and $\ee \ra \ee H$. The particle is relatively easy to detect
especially in the $ZH$ channel, where the main background from $ZZ$ pair
production can be suppressed efficiently by using micro--vertex detectors since
Higgs bosons with masses below 140 GeV decay mainly into $b\bar{b}$ pairs. Once
the Higgs boson is found, its fundamental properties can be investigated. The
Higgs spin can be measured by analyzing the angular dependence of the $ZH$
production process and of the Higgs decays into massive gauge bosons. The Higgs
couplings to the massive gauge bosons can be determined through the production
rates, the coupling to heavy fermions through the Higgs decay branching
fractions, and in some mass window, Higgs radiation off top quarks.

An even stronger case for $\ee$ colliders operating in the 500 GeV range is
made by supersymmetric extensions of the Standard Model. Since in the minimal
extension, the lightest Higgs particle has a mass below 140 GeV and decays
mainly into $b\bar{b}$ and $\tau^+ \tau^-$ pairs, it cannot be missed at an
$\ee$ collider with an energy $\sqrt{s} > 300$ GeV, independently of its decay
modes and in the entire $\SUSY$ parameter space. The heavy neutral Higgs
particles can be produced in the bremsstrahlung and fusion processes or
pairwise, these processes being complementary.  At least one neutral Higgs
boson must be detected, and in a large part of the $\SUSY$ parameter space, all
neutral Higgs bosons can be observed. Charged Higgs particles can be detected
up to
practically the kinematical limit.

Future $e^+e^-$ can be turned to very high--energy $\gamma \gamma$ or $e
\gamma$ colliders by using back--scattering of laser light. The $\gamma \gamma$
mode of the $e^+ e^-$ collider could be useful to measure accurately the
Higgs--photons coupling to which new particles might contribute, and to study
the ${\cal CP}$ properties of the Higgs particles.

$\ee$ linear colliders operating in the 300--500 GeV energy range and
hadron colliders operating in the multi--TeV range have a complementary
potential for adressing the key issue of the mechanism of electroweak symmetry
breaking.

\nonumsection{Acknowledgements}

\nn Thanks go first of all to Fernand Renard and Peter Zerwas. Some of the work
presented in this review has been elaborated jointly with Peter Zerwas; I thank
him for an enriching, exciting and fruitful collaboration. I also thank my
other main collaborators on this subject: Michael Spira [also for his help in
preparing some of the figures of section 3], Paolo Gambino and Jan Kalinowski.
Fernand Renard, Jean--Loic Kneur and the members of the Laboratoire de Physique
Mathematique of Montpellier University have provided a hospitable working
environment during this winter; I would like to express my gratitude to them.
I also would like to thank Profs. G. Auberson, J. J. Aubert, P. Binetruy, P.
Chiappetta, P. Fayet, G. Mennessier and F. M. Renard, for accepting to
referee my habilitation thesis. Finally, thanks to Manuel Drees for many
useful critical remarks on the manuscript. This work is supported
by the Natural Sciences and Engineering Research Council of Canada and by the
French Centre National de la Recherche Scientifique. \s

\nonumsection{References}

\end{document}